\documentstyle[twocolumn,prd,aps]{revtex}
\begin{document}
\draft
\title{\bf 
The search for   ``polarized'' instantons in the vacuum.}
\author{M.~Yu.~Kuchiev\cite{byline}}
\address{School of Physics, University of New South Wales,
Sydney, 2052, Australia. }
\date{19 February 1996}
\maketitle
\begin{abstract}
The new phase of a gauge theory 
in which the instantons are ``polarized'', i.e.
have the preferred orientation is discussed.
A class of  gauge theories with the specific condensates of the scalar 
fields is considered.
In these models there exists an  interaction 
between instantons   resulting from one-fermion loop
correction. The interaction makes the identical
orientation of instantons to be the most probable,
permitting one to expect the system to  undergo
the phase transition into the  state with polarized  instantons.
The existence of this phase  is confirmed in the mean-field
approximation in which there is
the first order phase transition separating the ``polarized phase''
from the usual non-polarized one.
The considered phase  can  be important for 
the  description of gravity in the framework of the gauge field theory.
\end{abstract}


\section{Introduction}
%*******************************************************************
In this paper the new nontrivial vacuum state of  gauge theory is
discussed.
In this state instantons are ``polarized'', 
i.e. have the preferred orientation.  This polarized
state is  interesting by
itself, but its importance is  enhanced by  Ref.\cite{IAP} where 
a scenario was suggested 
to describe effects of gravity in the framework of 
Yang-Mills gauge theory.
In this approach
space-time is  supposed to be  basically flat,
there are only  usual for the Yang-Mills theory 
fields of spin $0, 1/2$  and $1$ on  the 
basic level of  the theory. 
There are neither 
gravitons  nor the dimensional Newton gravitational constant  in the
basic Lagrangian of the theory. 
In spite of its ``conventional nature"  the gauge theory as discussed in 
\cite{IAP} can provide a description of the effect of gravity. 
In recent decades  it was supposed 
that   quite new basic physical conceptions, such as
 supergravity, strings and superstrings,  for a review see  \cite{Br},
are vital for quantum gravity.

In Ref.\cite{IAP} 
a  specific phenomenon 
based upon nontrivial 
topological excitations of the gauge field was considered.
The most clear  topological excitation is the instanton \cite{Bel}.
It  has several degrees of freedom: its position, radius and  orientation. 
Generally speaking an instanton  may be oriented arbitrarily.  
It was assumed in \cite{IAP} that 
it is possible to construct such the Yang-Mills theory
that the instantons in the vacuum state have
a preferred direction of their 
orientation.
That means that the probability for any instanton to be oriented along
certain direction is greater than along any other direction.
This phenomenon can be called  ``a polarization'' of  instantons.
 For formulation of the problem discussed
in this paper it is important 
to recall more  precisely the main result of \cite{IAP}. 
With this purpose
consider $SO(4)$ gauge theory. Then the  gauge algebra $so(4)$
is the sum of two $su(2)$  subalgebras: $so(4)=su(2)+su(2)$.
It was assumed in \cite{IAP} that the vacuum state of $SO(4)$ gauge theory
possesses the following properties.
First, instantons belonging to one $su(2)$ subalgebra of the 
gauge  algebra are polarized.
At the same time the antiinstantons belonging
to this $su(2)$ subalgebra are not polarized. Second, 
there is the reversed situation in the other $su(2)$ subalgebra:
the antiinstantons belonging to that subalgebra are
polarized while the instantons of that subalgebra are not polarized.
In other words,  in both $su(2)$ subalgebras the situation
is nontrivial: in one of them the instantons are
 polarized  while in the other one  the antiinstantons
are polarized. Note however, that the topological charge in every one
$su(2)$ subalgebra might be zero resulting in equal
 concentrations of instantons and antiinstantons for the given 
subalgebra.

The  concentrations of  these polarized instantons of
one $su(2)$ subalgebra and polarized antiinstantons
of the other $su(2)$ subalgebra are supposed to be finite and equal.
One can call this phenomena the
 ``condensate of polarized instantons and antiinstantons''
or ``Instanton-Antiinstanton Polarization'' in the vacuum.
 Certainly this desired
vacuum state can not be achieved in the framework of the pure
gauge theory. The theory must acquire some special properties
in the scalar and fermion sectors as discussed in detail below.
The main result of Ref.\cite{IAP} is that if the vacuum of the 
$SO(4)$ gauge theory  possesses the condensate discussed then
this gauge theory  reveals  the effect of  gravity. 

Note first of all 
that  existence of the considered nontrivial phase of the gauge field 
 does not come into contradiction
with the gauge invariance of the theory.
The gauge invariance forbids those nontrivial phases whose order parameter
is not invariant under local gauge transformations \cite{Elit},  
\cite{Creu}, \cite{Polya}.
The orientation of instantons 
cannot be varied by the
$local$ gauge transformations. This is the well known fact for 
the pure gauge theory, see Refs.\cite{Brown},\cite{ChW}.
% It can be easily generalized for the case when there are
% the scalars developing the scalar
% condensates and fermions in the theory, as is the case considered
% in this paper. For our purposes it is sufficient  to consider  the case
% of  instantons so small that their individual basic properties  are  not 
% affected by the considered scalar condensates.
% This makes the orientations of instantons to remain invariant under
% local gauge transformations for the theory with 
% scalars and fermions as well.
Therefore the instanton orientation may play the role
of an  order parameter for the considered 
nontrivial phase of the gauge theory.

The most ``natural" way to look for the condensation of  polarized instantons
is to find  an  interaction between instantons
which forces any two instantons to have the identical orientation.
Then one can
expect  the system to undergo a  phase transition  
into the state with polarized instantons.
An attempt to implement this idea  meets two problems.
First, in a pure gauge theory 
instantons do not interact. 
There is the well-known interaction between instantons and antiinstantons 
\cite{CDG},
but the interaction between instantons seems to be questionable because there 
is an  exact multi-instanton solution  with arbitrary
orientation of instantons  \cite{ADHM}. 
Second, along with instanton interaction there must be
no interaction between antiinstantons in order to keep
the system of antiinstantons disordered, as it is vital
for the considered above construction of Ref.\cite{IAP}.

The  models considered in this paper
 include the scalars and fermions 
in  a special way resolving  the above mentioned problems. 
The simplest model is based upon
 the $SU(2)$ gauge group. The models  provide
 the desirable effective interaction between  
instantons  making  their
identical orientation   the most probable. 
This result gives a hope that the phase transition into the  state 
with polarized instantons is possible.
Moreover, the mean field approximation confirms the existence of the
phase with polarized instantons. It is separated from the non-polarized
phase by the first-order phase transition.

It is important
that the models considered provide
the interaction of instantons
belonging to  the 
$su(2)$ gauge algebra, but no interaction
between the antiinstantons of this algebra.
As a result one can expect
 no condensation of antiinstantons
belonging to this $su(2)$ gauge algebra. 
The models can describe the  reversed situation: for a given $SU(2)$ gauge 
group they can provide the
 antiinstantons with the interaction while
instantons would possess no interaction. Then in this $SU(2)$
gauge group one can expect the
antiinstantons to become polarized, while instantons remain non-polarized.
This is exactly what is necessary
for the  discussed above $SO(4)$ gauge group construction, where instantons
must be polarized for one $su(2)$ subalgebra and antiinstantons -
for the other.

Notice that for $SO(4)$ gauge group the necessary
vacuum state of polarized  instantons and antiinstantons 
does not violate the parity conservation law in spite of the fact that 
there is a clear distinction between instantons and antiinstanton.
The polarized instantons  
belong to the $su(2)$ subalgebra 1, the antiinstantons to the
$su(2)$ subalgebra 2.
The inversion  results in the transformation 
$P$:   instantons $\rightarrow$
antiinstantons, and antiinstantons
$\rightarrow$ instantons. This, however, does not produce the new vacuum 
state.
Really, there is a freedom to arbitrarily choose the two $su(2)$
subalgebras. Simultaneously with inversion  one
can change the names of the subalgebras, $P$: subalgebra 1 $\rightarrow$
subalgebra 2, subalgebra 2 $\rightarrow$ subalgebra 1. Under this 
transformation
 the polarized instantons remain in the subalgebra 1,
the antiinstantons in the subalgebra 2. 
In the new vacuum state the orientations of instantons and antiinstantons
differ from orientations in the initial vacuum. 
In order to restore the initial orientation it is sufficient to
fulfill the global rotation in the isotopic space, which is allowed due to
gauge invariance of the theory.
The vacuum state remains invariant under thus understood
inversion. If all the other properties of the theory are
invariant under the transformation:  subalgebra 1 $\rightarrow$
subalgebra 2, subalgebra 2 $\rightarrow$ subalgebra 1,
then the theory satisfies the parity conservation law. 
In this paper we deal mainly with  $SU(2)$ gauge group. For this
gauge group we are looking for the possibility to
construct the vacuum with polarized instantons (or antiinstantons).
The  single $SU(2)$ group with polarized instantons
obviously violates the parity conservation law, 
because inversion results in the transformation of the
instantons into antiinstantons. We will 
keep in mind that the parity would be restored, if one combines
two $SU(2)$ groups, having polarized instantons in one of them and 
polarized antiinstantons in the other.

Consider the main idea.
The usual way to govern the properties of the gauge theory is 
provided by the scalar condensate. It is clear that this
condensate itself cannot
resolve the puzzle of the different behavior of
instantons and antiinstantons. In order to do this let us remember
that the instantons and antiinstantons interact differently
with right-hand and left-hand fermions.
This property manifests itself most strongly for fermion zero-modes:
the instantons produce the  right-hand fermionic zero-modes,
while the antiinstantons give the left-hand zero-modes \cite{tH}.
Realizing that we can develop the following construction.
Let there exist some condensates of the scalars. Let the scalar-fermion
interaction has the scalar-pseudoscalar vertex proportional to $1-\gamma_5$.
Then the right-hand fermions interact with the scalar condensate,
while the left-hand fermions do not. As a result the right-hand
zero modes of 
fermions provide a connection between the scalars and instantons.
It  permits the scalars to influence upon the instantons in a 
specific manner.
We will see that this influence results in the desirable interaction 
between the instantons.  
No similar connection  appears
between scalars and the  antiinstantons  because the left-hand fermions
do not interact with the scalar condensates.

This construction may be described in the usual terms as a one-fermion-loop
correction  to the gauge field action. 
Calculating it we are
to consider the interaction of the fermions with the scalar condensate
as well as with the gauge field created by several instantons.
In order to find the necessary effect   
one must choose properly the scalar condensates. 
The condensates create the field $V$ applied to the fermions.
It is clear that 
this field must not commute with the gauge field created by instantons,
$[\gamma \nabla , V]\ne 0$, 
where $\nabla$ is the
covariant derivative $\gamma \nabla=
\gamma_\mu(\partial_\mu  - i A^a_\mu(x)T^a$), $T^a=\tau^a/2,~a=1,2,3$
are the generators of the $SU(2)$
gauge group, and $A^a_\mu(x)$ is the gauge field 
created by several instantons. Otherwise no connection between
the scalar condensates and instantons would appear. If one 
 wishes to consider homogeneous
 field $V$, as is usual,
then the only way to satisfy this condition is
to suppose that the field $V$ depends on the generators of the gauge group:
$V \sim \vec T\vec U(1-\gamma_5)$.  As a  result we are to
introduce into the problem some additional vector $\vec U$ to multiply
the vector of generators $\vec T$. 
The  constant vector $\vec U$
not only looks ugly but makes no good, as can be 
verified. 
We are to find the vector  whose averaged value is zero, but the averaged
values of its powers could play a role: $<U^a>=0,~<U^a U^b>=(1/3)
\vec U^2\delta_
{ab}, ~...~$. The way to do this can only  be provided by an  additional 
symmetry.
We are  to consider some ``additional'' $SU(2)$ group and identify
its generators with the necessary vector $\vec U$.
There are two ways to introduce this additional  symmetry.
It may be considered either as a global symmetry or as a local gauge
symmetry. For the first case one can visualize  group
as ``a flavor group'' or ``a group of generations''. 
This possibility was first discussed 
 in Ref.\cite{det}.
For the second
case the gauge group considered becomes as big as $SU(2)\times SU(2)$. 
Both possibilities and shown to
provide the system of instantons with the desired properties.
There is meanwhile a  substantial difference between these two
realizations of the model. For the global additional group
the gauge field acquire a mass. In contrast, the local realization
of the additional symmetry permits to keep the
gauge fields for some $SU(2)$ gauge subgroup 
massless. The latter property is 
desirable because there must remain the massless gauge field
if we wish to construct afterwards the massless gravitons with the help 
of this field.

Note that there is a clear and  interesting
analogy between the discussed models and the 
phenomenon of ferromagnetism. 
The  problem of the phase transition into a state
with polarized instantons resembles the  transition into
a ferromagnetic phase.
The instantons play a role very similar to the role of
magnetic impurities in the ferromagnetic case.
The fermionic zero modes resemble the atomic outer electrons.
In the pure gauge theory the fermionic zero-modes for several
instantons remain degenerate in accordance with the index theorem 
\cite{AS}, \cite{Kis}.
This resembles the situation when atomic outer electrons are
well-localized on the atoms and their energy levels are degenerate.
The interaction of right-hand zero-modes with the scalar 
condensates results in
the splitting of the  zero modes. 
Therefore these scalar condensates play
the same  role as  
the crystal field which results in the splitting of the atomic energy levels
and creation of a conducting band. 
The splitting  makes zero-modes look  similar to the 
electrons in the conducting band.
The problem of  instanton interaction looks
very close to the origin of the exchange integral 
describing interaction of the magnetic impurities
in the  ferromagnetic theory.
The exchange integral between impurities
appears due to  their interaction with
the electrons of the conducting zone.
Similarly the interaction of instantons with the right-hand zero-modes
results in the effective interaction between instantons.
If some other impurities do not interact with the electrons of the conducting
zone then they do not play a role in the ferromagnetic state. Similarly,
the antiinstantons which do not interact with the right-hand
fermions  do not play a role in the considered problem.

The precise results of this paper are the following. 
The fermionic determinant $\det {(F)}$ is calculated
in  the one-loop approximation
when fermions interact with instantons as well as with
the scalar condensate.  If we define 
\begin{equation}\label{detF}
\det {(F)} = \exp (-S_F)~,
\end{equation}
then  $S_F$ may be considered as an effective action, i.e. 
a contribution of the fermions to the action for the gauge field.
It is found that it describes the interaction 
of the instantons.
For  the simplest case of two well separated instantons this interaction
is found to be 
\begin{equation}\label{SF}
S_F = \frac {f^2\phi ^2}{2 m^2} \frac {\rho_1^2 
\rho_2^2}{r^4} \sin^2\gamma~,
\end{equation}
where $ \phi$ is the value of the scalar 
condensate,
$f$ is the coupling constant of the scalar-fermion interaction,
$m$ is the mass of fermions,  $\rho_1, \rho_2$ are the radii of the 
instantons 
and $r$ is their separation, $\gamma$ is the angle between the directions
of instanton orientation.
It is assumed in (\ref{SF}) that $\rho_1,  \rho_2  \ll r$.
Formulas (\ref{detF}), (\ref{SF}) describe the interaction between the 
instantons.
The action (\ref{SF}), first 
reported in Ref.\cite{det}, has a desired minimum
at  $\gamma = 0$
for the identical orientation of instantons. 
Note that usually the radiative corrections renormalize 
 physical quantities. In the case considered  they
 provide the new phenomenon:
the interaction between the instantons. 

The mean field approximation for the ensemble of instantons
interacting via action Eq.(\ref{SF}) is considered.
This approach confirms the existence of the nontrivial
vacuum state with polarized instantons. It  reveals also 
the first-order phase transition  which
separates the polarized phase  from the non-polarized.

In Sections II the model based on $SU(2)$ gauge theory with
additional global  $SU(2)$ symmetry is considered.
The interaction of the instantons is considered in detail in Sections 
III-V. In Section VI the mean field approximation is developed.
In Section VII the $SU(2)\times SU(2)$ gauge theory model is discussed.
Throughout the paper we use the set of definitions presented
in Section IX. The vital properties of the fermionic zero-modes
in the field of several instantons 
and their role in the instanton-instanton interaction is discussed
Sections X,XI.

\section{the $SU(2)$ model}
%*******************************************************************
Consider the $SU(2)$ gauge theory. Suppose that there are two generations
of fermionic fields in the fundamental representation
of this  gauge group. The masses
of the fermions are supposed to be equal and we will treat them as a 
 doublet in the space of generations. 
Suppose also that there are scalars in
the vector representation of
this gauge group. Consider
three generations of scalars, e.g. a scalar triplet in
the space of generations.
Let us introduce the interaction between the scalars and the right-hand 
fermions described by the Lagrangian 
\begin{equation}\label{Lscalferm}
{\cal L} _{sf} (x)=  f  \psi_A^+(x)  \Phi_i(x) U^i_{AB} \frac{1-\gamma_5}{2}  
\psi_B(x)~.
\end{equation}
Here $ f$ is a dimensionless constant of scalar-fermion interaction, 
$\psi_A(x)$
is the fermion doublet, indexes $A,B = 1,2$ label the variables in the space 
of generations, $ \Phi_i(x) , i=1,2,3$ is the triplet of scalar fields.
There is a freedom of choosing the matrixes $ U^i = U^i_{AB},~ i=1,2,3$
describing the coupling between different generations of fermions and scalars.
We choose these matrixes to be  the  triplet of generators of  
rotations in the space of generations 
\begin{equation} \label{vecU}
 U^i = U^i_{AB} = \sigma^i_{AB}/2,~ i=1,2,3~.
\end{equation}
Note that Euclidean formulation is used,
see for example Ref.\cite{VZNS}.

Notice that the Lagrangian Eq.(\ref{Lscalferm}) obviously
violates parity conservation law. It can be restored if,
alongside with the considered $SU(2)$ gauge group, we consider the
other $SU(2)$ gauge group in which the Lagrangian has the  form
similar to Eq.(\ref{Lscalferm})  but with the positive 
sign in front of $\gamma_5$.

The scalar fields $\Phi _i(x)$ are in the vector representation 
 of the gauge $SU(2)$ group, therefore
\begin{equation}\label{Phiai}
\Phi _i(x)=\Phi _{i,a}(x)T^a~,
\end{equation}
where $T^a=\tau^a/2,~  a=1,2,3$ are the generators of $SU(2)$ 
gauge transformations. 
Thus we have nine scalar fields $\Phi _{i,a}(x)$.
Suppose now that  their nonlinear self-interaction  results in 
developing of the
scalar condensate which has the  following form
\begin{equation}\label{scalcond}
 (\Phi _{i,a}(x))_{cond} = \phi \delta_{ia}~,
\end{equation}
where $\phi$ is a constant. 
One can always fulfill this condition choosing 
appropriately the self-interaction
of the scalars.
Then it follows from 
(\ref{Lscalferm}),(\ref{scalcond}) that  there appears the field $V$,
 \begin{equation}\label{V}
V=  f \phi~ (\vec T  \vec U) ~\frac{1-\gamma_5}{2} ~,
\end{equation}
which  influence upon the right-hand fermions.

Our goal is to calculate the
fermionic determinant $\det{(-i\gamma \nabla -i m -i V)}$.
It depends on the gauge field $A^a_\mu(x)$, 
which stands in the covariant derivative 
\begin{displaymath}
\gamma \nabla=\gamma_\mu \nabla_\mu= 
\gamma_\mu \left(\partial_\mu  - i A^a_\mu(x)T^a\right)~.
\end{displaymath}
We wish to evaluate the determinant when the gauge field is created by
several instantons. 
The gauge field and the field $V$ created by the scalar condensate
do not commute $[\gamma \nabla, V]\ne 0$, as follows from Eq.(\ref{V}).
This makes the determinant to depend non-trivially  on 
the field $V$  (\ref{V}).
The  determinant is 
calculated below in the one-loop approximation. This means that
the gauge field and the scalar condensate field are
considered as the given external fields. 

Let us present the fermionic determinant as
\begin{eqnarray} \label{detdet}
&&\det {(-i\gamma \nabla  -i m -i V)}=
\\
&&\det {(-i\gamma  \nabla  -i m )}\det{(1+GV)}~,
\end{eqnarray}
where $G$ denotes the propagator 
describing  behavior of the massive fermions in the  gauge field 
\begin{equation}\label{G}
G=(\gamma \nabla  + m)^{-1}~.
\end{equation}
The first factor $\det {(-i\gamma \nabla -i m )}$ in (\ref{detdet})
describes the known fermion behavior in the pure gauge field.
It is not relevant to the effect considered.
The important
information is contained in the  second factor 
\begin{equation}\label{detGV}
\det{(F)}=\det{(1+GV)}=\exp{( -S_F)}~.
\end{equation}
which we will discuss in detail. 
The determinant $\det{(F)}$   
may be considered as a contribution of the fermions to the action
for the  gauge field $S_F$.

The instantons are known to
create the zero-modes of the fermions. These zero-modes  play a
crucial role in the following calculations. Therefore it is useful
to distinguish them in the fermion propagator. 
With this purpose let us introduce
the projection operator $P$ onto the states of zero modes. 
It satisfies the conditions
\begin{equation}\label{P1}
P^2=P~, ~(\gamma \nabla ) P = 0,~\mbox{Sp}(P)=2 k~.
\end{equation}
Here $k$ is the number of  
instantons. The number of zero-modes in the considered case of
two fermionic generations in the fundamental representation
of the $SU(2)$ group is $2 k$.
The propagator (\ref{G})
may be presented as
\begin{eqnarray}\label{G01}
G~&=&G_0 + G_1,\\  \label{G0}
G_0&=&P/m,\\ \label{G1}
G_1&=&(1-P) (\gamma  \nabla +m)^{-1}(1-P).
\end{eqnarray}
To simplify  calculations let us consider the instantons  so small 
 that the condition
\begin{equation}\label{mrho}
m \rho \ll 1~,
\end{equation}
where $\rho$ is the  instanton radius is fulfilled. Then the fermionic mass
$m$ may be considered as
a small parameter and we will put $m=0$ wherever it is possible. 
Notice that the instanton separation is not restricted in this consideration.
We will return to this point later, see Section V.A.

The propagator of the   nonzero-modes for massless case is simplified to be
\begin{equation}\label{G1simpl}
G_1^{(m=0)}=(1-P) (\gamma \nabla  )^{-1} (1-P).
\end{equation}
This propagator may be presented in the convenient form
 found in Ref.\cite{Brown}
\begin{equation}\label{G1simpBr}
G_1^{(m=0)}=(\gamma \nabla)\frac{1}{\nabla^2}
\frac{1+\gamma_5}{2}+\frac{1}{\nabla^2}
(\gamma \nabla)\frac{1-\gamma_5}{2}~.
\end{equation}
It is clear from (\ref{V}),(\ref{G1simpBr}) that
\begin{equation}\label{GVG}
VG_1^{(m=0)}V=0~.
\end{equation}

Using Eq.(\ref{GVG}) we find that only the zero-modes (\ref{G0}) 
give a contribution
to (\ref{detGV}), while nonzero-modes are eliminated
\begin{equation}\label{zerodet}
\det{(F)}=\det{\left( 1+\frac{1}{m}PV\right) }~.
\end{equation}
This result greatly simplifies the problem 
because for the finite number of the instantons
the operator $PV$ is presented by a finite matrix.

It is necessary to keep in mind that the
scalar condensate (\ref{scalcond}) results in the creation of the mass
for the gauge field through the Higgs mechanism \cite{Higgs}
\begin{equation}\label{mV}
M_V^2=\frac{3 }{4}g^2 \phi^2~,
\end{equation}
where $g$ is the gauge coupling constant.
We will neglect this mass supposing that
the instantons radii $\rho$ 
are sufficiently small
\begin{equation}\label{restMV}
\rho M_V  \ll 1~.
\end{equation}

\section{ The  fermionic determinant}
%*******************************************************************
In order to  calculate the determinant (\ref{zerodet}) we are
to consider the matrix $V$ in the subspace of zero-modes where it has a
form
$V=\langle t,B | V | s,A\rangle $, here the indexes $s,t=1,\cdots,k$ numerate 
the zero-modes in one
generation of instantons, while the indexes $A,B=1,2$ label the generations
of zero-modes.
According to (\ref{Psixan}) 
 the wave-functions of the zero-modes have the form
\begin{eqnarray}\label{wf}
\Psi_{s,A}=\Psi_{s,A}(x,\alpha,n)=\frac{1}{\pi}
[L^+_s(x)]_{n, n'}\epsilon_{n',\alpha} \omega_A~.
\end{eqnarray}
Here the index
$\alpha=1,2$ is a spinor index of the right-hand spinors, 
the index $n=1,2$ is the isospinor index,
$\epsilon_{\alpha, n}$ is the usual $2\times 2$ antisymmetric tensor.
$L^+_s(x)$ is a quaternion  conjugate to
$L_s(x)$, defined in Eq.(\ref{Ldef}).
According to this definition 
$[L^+_s(x)]_{n, n'}=L_{s,\mu}(x)(\tau^-_\mu)_{n,n'}$.
The function $\omega_A$ in (\ref{wf})
describes the  orientation of a zero-mode in the two-dimensional
space of fermionic generations,  it is normalized as  
$\langle \omega^+_B |\omega_A\rangle =\delta_{AB}$. 

Using  the wave-functions (\ref{wf}) we
find the matrix of the operator $V$ (\ref{V})
\begin{eqnarray}\nonumber
\langle t,B | V | s,A\rangle =\sum_{\alpha,n,n'}\int d^4x 
&&\Psi^+_{t,B}(x,\alpha,n') 
(n',B |V|n,A) \\ \label{matV} && \Psi_{s,A}(x,\alpha,n)~,
\end{eqnarray}
where 
\begin{equation}\label{whe}
( n',B |V|n,A) =\frac{f\phi}{2} \vec \tau_{n',n} \vec U_{BA}~.
\end{equation}
After simple algebraic transformations we evaluate
from (\ref{wf}),(\ref{matV}),(\ref{whe}) the  convenient form for the 
matrix 
\begin{equation}\label{matrix}
\langle t,B | V | s,A\rangle =-i\frac{f\phi}{2}  \vec W_{t,s} \vec U_{BA}~,
\end{equation}
where
\begin{equation}\label{ttaus}
\vec W_{t,s}=\frac{2}{\pi^2}\mbox{Re}\int L_t(x) i\vec \tau L^+_s(x)~ d^4x~,
\end{equation}
% =  \frac{2}{\pi^2}\{\mbox{Im}\int  L^+_s(x) \tau L_t(x)~ d^4x~.
$s,t=1,\cdots,k$.
According to (\ref{ttaus}) the matrix $\vec W_{s,t}$ is 
antisymmetric 
\begin{equation}\label{zero}
\vec W_{t,s}=-\vec W_{t,s}~.
\end{equation}
Using (\ref{matrix}) we can find the explicit 
expression for the determinant (\ref{zerodet}) for a small number $k$ of 
instantons.
For a trivial case of single instanton $k=1$ we
 find $\det{(F)}=1$,  due to Eq.(\ref{zero}).
For two instantons, $k=2$, we find from Eq.(\ref{matrix})
\begin{equation}\label{W12}
PV = \frac{f \phi}{2}
\left(  \begin{array}{cc}
\hat 0& -i   \vec W_{1,2} \vec U\\
i   \vec W_{1,2} \vec U &\hat 0
\end {array}\right)~.
\end{equation}
Remember that $\vec U$ is the $2\times2$ matrix in the space of generations.
In Eq.(\ref{W12}) $\hat 0$ is the zero $2\times2$ matrix in the same space.
Diagonalize the matrix in Eq.(\ref{W12})  we find the determinant
Eq.(\ref{zerodet})
\begin{equation}\label{detk=2}
\det{(F)}=\left( 1-\frac{1}{4}\zeta^2  \vec W^2_{1,2}\right) ^2~.
\end{equation}
% !REMEMBER. The coefficient $1/4$ here comes from the square of the
%  operator $U$ which is $\vec U = \vec \sigma/2$, compare $n^2$ in 
%  (\ref{arbspin}).
Parameter $\zeta $ is
\begin{equation}\label{zeta}
\zeta =\frac{f \phi}{2 m}~.
\end{equation}
% ! REMEMBER. The coefficient $2$ in denominator in the last equation comes 
%   from the generator of gauge transformations $\vec T=\vec \sigma/2$.

It is instructive to look at Eq.(\ref{detk=2}) from the point of view of
the eigenvalue problem for the operator $\gamma \nabla +V$.
 In the absence of the perturbation there are
four zero modes, which are the solutions of $\gamma \nabla \Psi =0$ with
zero eigenvalue. The perturbation $V$ results in the shift of the 
eigenvalues from the zero value
\begin{equation}\label{eigen}
\left( \gamma \nabla + V\right) \psi = \varepsilon \psi~.
\end{equation}
We restrict our consideration to the subspace of the zero modes,
where the operator $\gamma \nabla+V$ is identical to $V$,
$\gamma \nabla+V \equiv PV$,  given in  Eq.(\ref{W12}).
The  four-fold degenerate zero eigenvalue
is  split into two two-fold degenerate eigenvalues 
$\varepsilon_+$ and $\varepsilon_-$,
\begin{equation}\label{epspm}
\varepsilon_\pm = \pm \frac{f \phi}{4} |\vec W_{1,2}|~.
\end{equation}
Using them one finds the determinant $\det(F)=(1-\varepsilon_+/m)^2
(1-\varepsilon_-/m)^2$ which, of cause, is identical to Eq.(\ref{detk=2}).
This consideration
clarifies the physical picture discussed in the introduction:
the considered fermionic determinant appears
 due to  splitting of the zero modes. 

%Using the  same methods   we find for three instantons, $k=3$,
%\begin{eqnarray}\nonumber
%&&\det{(F)}=\biggl( 1-\frac{1}{4} \zeta^2 ( \vec W^2_{1,2}
%+ \vec W^2_{1,3}+ \vec W^2_{1,3}) +
%\\ \label{detk=3}
%&&\frac{1}{4} \zeta^3 \vec W_{1,2}( \vec W_{2,3} \times 
%\vec W_{3,1} )\biggr) ^2~.
%\end{eqnarray}
%The general formulas for   $\det{(F)}$ are more complicated if
%the number of instantons is bigger, $4\le k$.

An important case provides the perturbation theory 
limit $\zeta^2 \vec W_{t,s}^2 \ll1$.
We find from (\ref{detk=2}),(\ref {detGV})
\begin{equation}\label{perth}
S_F=\frac{\zeta^2}{2} \vec W_{1,2}^2~. 
\end{equation}
%The generalization of this result to the case of arbitrary
%number of instantons $k$ is strait-forward:
%$S_F= (\zeta^2/2) \sum_{s < t}  \vec W_{t,s}^2$.
%Up to now we
%considered two generations of fermions. 
%That was done for the sake of simplicity,
%we can consider the higher multiplets in a similar way.
%Suppose that there are $2 p+1$ generations of fermions in the fundamental
%representation of the $SU(2)$ gauge group. Let the matrixes 
%$\vec U$ in (\ref{Lscalferm})
%represent the generators of the rotation in the space of generations
%for the considered  $2 p+1$ multiplet. Consider again the most simple case
%of two instantons $k=2$.  
%It is easy
%to verify that representation (\ref{W12}) for the operator $1+PV/m$
%remains valid if we consider $\hat 1$ and $\vec U$ as $(2p+1)\times (2p+1)$
%matrixes in the space of generations. 
%Diagonalization of this matrix gives
%\begin{equation}\label{arbspin}
%\det{(F)} = 
%\prod_{0 \le n \le p}\left( 1-n^2\zeta^2  \vec W^2_{1,2}\right) ^2~.
%\end{equation}
%Here $n=p,p-1,\cdots$.
%For one generation  $p=0$ and
%(\ref{arbspin}) gives  the trivial result $\det{(F)} =1$.
%For two generations of fermions we have $p=1/2$ and
%(\ref{arbspin}) reproduces (\ref{detk=2}).

The presented results show that the fermionic determinant depends on the
 matrix elements $\vec W_{t,s}$ given in Eq.(\ref{ttaus}). If all 
instantons have identical orientation $q_t=\rho_t w,~t=1,\cdots,k$,
where $\rho_t$ is radius of $t$-th instanton,  and $w$ satisfying 
condition $w^+w=1$ describes the instantons orientation,
 then the matrix elements
vanish  $\vec W_{t,s}=0$. This follows from  representation (\ref{Ldef})
for the functions $L_s(x)$ given in Section VIII, 
but can be easily understood without calculations. For identical
orientation there is no ``complex'' parameter in the problem.
This eliminates the real part of the 
matrix element from the ``complex'' operator $i\vec\tau$
in Eq.(\ref{ttaus}).
As a result the determinant in this case 
is trivial $\det(F)=1$. We will see below that 
in  general the determinant is the function
of the instanton orientations.

\section{Large separation of two instantons} 
%*******************************************************************
Consider the situation when the radii of the two instantons are less then
their separation,  and therefore the dilute gas approximation is valid.
Let us calculate explicitly the fermionic determinant in this case.
According to Eq.(\ref{matrix}) we are to calculate the matrix element
$\vec W_{1,2}$ defined in Eq.(\ref{ttaus}). 
The functions $L_s(x)$ in the dilute gas approximation
are given in (\ref{Lgas}).
Substituting them in Eq.(\ref{ttaus}) we find
\begin{equation} \label{intLWL}
\vec W_{1,2}=\frac{2}{\pi^2}\mbox{Re} \int
\frac{x-y_1}{|x-y_1|^4}
q^+_1i\vec \tau q_2 \frac{x^+-y^+_2}{|x-y_2|^4} d^4x~.
 \end{equation}
The main contribution to the integral in the above expression 
comes from the region $|x-y_1|\sim |x-y_2|
\sim |y_2-y_1| $ which justifies inequality (\ref{dgcond}) used to evaluate
 $L_s(x)$ (\ref{Lgas}). The simple form obtained for the matrix
element (\ref{intLWL}) is due to the chosen 
singular gauge which makes the vector
potential to be strongly localized on the instantons, see Eq.(\ref{Adg}).
For our purposes it is  sufficient
to take the wave functions approximated by (\ref{Lgas}) in spite of the fact
that they
satisfy the orthogonal condition
(\ref{ReKK}) with accuracy $\sim 1/r_{12}^2$, which
is comparable with the right-hand side of (\ref{intLWL}).
This disadvantage does not manifest itself 
 because we calculate in (\ref{intLWL}) the matrix element
of the ``imaginary'' operator $i \vec \tau$ while in the orthogonal condition
(\ref{ReKK}) the matrix element from ``real'' unity operator is calculated,
see numerical results presented below in Fig.1.

Calculating the integral in (\ref{intLWL}) we find
\begin{equation}\label{wint}
\vec W_{1,2}=\frac{1}{r^2_{12}} (\delta_{\mu\nu}-
2v_\mu v_\nu) \mbox{Re} (\tau^+_\mu q^+_1  i \vec \tau q_2 \tau^-_\nu)~,
\end{equation}
where $r_{12}=|y_2-y_1|$ and $ v_\mu=(y_2-y_1)/|y_2-y_1|$. 
Taking into account identities
$\tau^+_\mu q \tau^-_\mu =4 \mbox{Re}(q)$ and $ \mbox{Re}( v q v^+)=
|v|^2\mbox{Re}(q)$
 valid for arbitrary quaternions $q,v$ we find
\begin{eqnarray}\label{Wfin}
\vec W_{1,2}&=&\frac{2}{r_{12}^2} \mbox{Re}(q^+_1 i\vec \tau q_2)~,
\\ \label{alpha}
\vec W_{1,2}^2&=&\frac{4}{r_{12}^4}|q_1|^2 |q_2|^2 \sin^2\gamma_{12}~,
\end{eqnarray}
where $\gamma_{12}$ is the angle beween the directions of the orientation of
instantons $1$ and $2$ defined by the identity $\mbox{Re}(q_1^+ q_2)=
|q_1| |q_2| \cos \gamma_{12}$. Remember that this angle is defined
up to mod $\pi$ because the instantons described by quaternions
$q$ and $-q$ give the same field, see Eq.(\ref{1ins}).

Substituting Eq.(\ref{alpha}) 
into Eq.(\ref{detk=2}) we find the determinant for two
well separated instantons
\begin{equation} \label{detk=2g}
\det(F)=\left( 1-\zeta^2 \frac{|q_1|^2|q_2|^2}{r_{12}^4}\sin^2 \gamma_{12}
\right)^2~.
\end{equation}
We are mainly interested in the region of large separation
between  instantons. If it is so large that  
  $\zeta^2 \vec W_{t,s}^2 \ll1$,  then we can consider the
perturbation theory limit 
(\ref{perth}).
Using (\ref{alpha}) we find 
the effective action
describing the interaction of $2$ instantons
\begin{equation}\label{SFdilgas}
S_F \approx S^0_F=2 \zeta^2 
\frac{|q_1|^2|q_2|^2}{r_{12}^4}\sin^2{\gamma_{12}}~.
\end{equation}
This formula reproduces Eq.(\ref{SF}) proclaimed in the 
introduction. The generalization of this result to the case of
several instantons is straightforward, $S^0_F= \zeta^2 \sum_{s\ne t}
|q_s|^2|q_t|^2\sin^2{\gamma_{st}}/r_{st}^4$.
%  REMEMBER.   Coefficients are: 
%  $1/4$       from $U^2$, see (\ref{detk=2})
%  $2$         from power $2$ in (\ref{detk=2})
%  $4$         from $W^2$ given by (\ref{Wfin}),(\ref{alpha}).
%                               Total $2$.

%We evaluated Eqs.(\ref{SFdilgas}),(\ref{SF})
%for the case of the doublet of fermion 
%generations,
%but from (\ref{arbspin}) it follows that they remain valid 
% for  arbitrary number $2p+1$ of  generations as well.
%The numerical coefficient $c$ in (\ref{SF})
%for arbitrary $p$
%is equal to
%\begin{equation}\label{coefc}
%c=c(p)=2 \sum_{0 \le n \le p} n^2=\frac{1}{3}p(p+1)(2p+1)~.
%\end{equation}

Note the peculiarity of $\det (F)$ (\ref{detk=2g}).
It possesses a node for specific values of 
parameters $\zeta, q_1,q_2,r_{12},\gamma_{12}$. 
If $\zeta>1$ then this node may be situated in the region of applicability
of the formula.
The node has a clear  physical meaning.
When it happens
then there is a ``compulsory''  creation of the four fermion-antifermion pairs
produced by the two considered instantons. To see this 
 consider the instanton contribution to the amplitude of the pair
creation. It has a multiple node coming from the determinant and the poles
produced by the Green functions describing the propagation of the created
 fermion pairs in the field of instantons.
 The node and the poles  compensate for each 
other resulting in the finite amplitude for the physical process.
 This effect is similar
to   creation  of
 a pair of massless quarks  in the field of an instanton in QCD, see for 
example \cite{VZNS}. The difference is that in the considered case the
fermions possess mass $m>0$. 
Note also that 
the fermionic determinant is
finite in the limit $m=0$.
The usual node of the determinant 
$\det(-i\gamma\nabla-im)$ in the pure gauge field at $m=0$, 
see Eq.(\ref{detdet}),
is compensated for by the pole arising from $\zeta$ in 
Eq.(\ref{detk=2g}).

Let us summarize the conditions under which the obtained results are valid.
The radii $\rho$ and separations of instantons $r$ must obviously
satisfy condition $\rho<r$ which justifies the dilute gas approximation.
Eq.(\ref{mrho}), $ m \rho < 1$,
guarantees the validity of the approximation based on consideration
of zero-modes only.
Eq.(\ref{SFdilgas}) was evaluated from Eq.(\ref{detk=2g})
in the perturbation theory limit $S^0_F<1$ valid if 
  the separation is large enough 
$(\zeta |\sin \gamma|)^{1/2}\rho<r$. 
The radii of instantons must be small enough
to protect them from  being cracked by the scalar condensate.
This means according to Eq.(\ref{restMV}) that $\rho < 1/(g\phi)$.

According to Eq.(\ref{detk=2g}) there is the interaction between
the instantons. 
Eq.(\ref{SFdilgas})  shows that 
it makes their identical orientation
$\gamma_{12}=0$ be the most probable. 
We did  not use  perturbation theory
 over the parameter
$\zeta=f \phi/(2m)$.
In this sense the problem of interaction between
instantons is solved exactly.

\section{The suppression of instanton interaction in the asymptotic regions}
It follows from from Eq.(\ref{SFdilgas}) that the effective action for two
 instantons depends on their separation as
$S_F \sim 1/r^4$. 
It is interesting
to consider the action averaged over the positions of instantons.
This averaged action has the 
logarithmic divergence $ (S_F)_{\rm av} \sim \int d^4r/r^4$ for
small separation $r=0$  as well as for large separation
$r =\infty$. 
The purpose of this section is to study what is happening 
 in the vicinity of $r\sim 0$ and $r\sim \infty$. 
It is shown that the interaction of instantons is 
suppressed in both these regions. 
 The origin of the suppression is different.
For small separation the two-instanton interaction itself is   
strongly suppressed when $r < \rho_1+\rho_2$. In contrast, there is no
suppression in the two-instanton problem for large 
separation where Eq.(\ref{SFdilgas}) gives a correct 
asymptotic result for the  two-instanton problem. 
The existence of other instantons changes this result drastically: the 
two-instanton interaction becomes suppressed exponentially in the presence
of the  other instantons.

\subsection{Small separation of two instantons}
Consider first interaction of the two instantons in 
the region of small separation $r_{12}\sim |q_1|,|q_2|$. 
Simple analytical results may be obtained in this region if we assume
that 
\begin{equation}\label{rl}
r^2_{12} \gg | q_1 q_2 \sin \gamma_{12} |~. 
\end{equation}
This inequality covers the interesting for us now case of small 
separation $r^2_{12}< | q_1 q_2|$
provided the orientation of the two instantons is close enough 
$|\gamma_{12}|\ll 1$. 
Remember that we are interesting in the ``polarized'' instantons, therefore
their identical orientation is of particular interest. Inequality (\ref{rl})
covers as well  the large separation region, which was considered above.
 It is shown in Section XI, see Eqs.(\ref{rinfgam}),
that  the matrix element describing interaction of the two
instantons satisfying Eq.(\ref{rl}) is
\begin{equation}\label{W12gam}
\vec W_{1,2}=  2 F(\kappa_1,\kappa_2)
\frac{ {\rm Re}(q_1^+ i \vec \tau q_2) }   {r_{12}^2} ~,
\end{equation}
Eq.(\ref{W12gam}) differs from Eq.(\ref{Wfin})
by the factor $F(\kappa_1,\kappa_2)$, where $\kappa_i = 2 |q_i|/r,~i=1,2$,
and  $F(\kappa_1,\kappa_2)$ is
defined  in Eq.(\ref{funuv}).

It is convenient to present the action Eq.(\ref{perth}) in the following form
\begin{equation}\label{f2as}
S_F = K S_F^0=
2  K \zeta^2 \frac{ |q_1|^2 |q_2|^2}{ r_{12}^4 } \sin^2 \gamma_{12}~.
\end{equation}
Here $K$ is the factor  which
shows how strongly the action $S_F$ deviates from the asymptotic
expression $S_F^0$  defined in Eq.(\ref{SFdilgas}). 
It follows from Eqs.(\ref{perth}),(\ref{W12gam})
that for the case under consideration this factor is 
$K = F^2(\kappa_1,\kappa_2)$.
Its behavior as a function of $r_{12}/(|q_1|+|q_2|)$ is illustrated in
Fig.1.
For large separation $r_{12}\gg  |q_1|+|q_2|$ the factor $K$ is close
to unity, as is should be in order to reproduce the large separation result
Eq.(\ref{SFdilgas}).
The interesting thing happens for small separation $r_{12} \rightarrow 0$. 
Here $K$ is decreasing very fast. 
It is easy to find from Eq.(\ref{funuv}) the asymptotic behavior
  $F^2(\kappa_1,\kappa_2)\rightarrow r_{12}^4 (\ln 1/r_{12} +{\rm const})^2,~
r_{12}\rightarrow 0$.
Fig.1 shows that this drastic
decrease happens starting  from $r_{12}\le |q_1|+|q_2|$.
Thus the interaction of the two instantons is  
strongly suppressed for small separation.
 According to Eq.(\ref{f2as}) 
the averaged action in the small-separation region
behaves as $(S_F)_{\rm av}\sim \int K d r/r$.
The later integral is  convergent, the main contribution to it
gives the region in which $r_{12}\ge |q_1|+|q_2|$. 
The other interesting feature
illustrated by Fig.1 is the fact that $F^2(\kappa_1,\kappa_2)$ 
is almost independent on the
ratio of the instanton radiuses $|q_1|/|q_2|$ thus depending
mainly on $r_{12}/( |q_1| + |q_2|)$.

Up to now we have discussed  the region covered by Eq.(\ref{rl}),
which gives the strong restriction on $\gamma_{12}$ for small separation
of instantons.
This was done in order to 
develop analytical calculations as far as it is possible.
The general case of small separation and arbitrary orientation 
of  two instantons can be handled  by direct  numerical calculations.
The useful   representation for
the matrix element $\vec W_{1,2}$ is given in Section XI in Eq.(\ref{Wconv}).
In  general case the factor $K$ defined by Eq.(\ref{f2as})
is a function of three variables
$K= K(\kappa_1,\kappa_2,\gamma_{12})$. 
The results of numerical calculations for $K$
based on Eqs.(\ref{Wconv}),(\ref{f2as}) are also  presented in Fig.1.
The suppression of  configurations
with the small separation $r_{12}\le |q_1|+|q_2|$ 
is the general feature, it is valid
for any orientations and radii of  instantons.
Notice the very smooth dependence of $K$ on $\gamma_{12}$.
It manifests itself mainly in the region of small separation, which 
is suppressed.
Thus the considered factor $K$ with reasonably good accuracy
may be considered as a function of only one variable 
$r_{12}/(|q_1|+|q_2|), ~
K \approx F^2\left( (|q_1|+|q_2|)/r_{12}, (|q_1|+|q_2|)/ r_{12} \right)$,
where the function $F \left ( \kappa_1,\kappa_2 \right )$ is defined in
Eq.(\ref{funuv}).

We conclude: the interaction of the instantons 
is  suppressed by the factor $K$ when their separation is small.

\subsection{Large separation of two instantons in the presence of other
instantons}
Consider two instantons  1 and 2 separated 
by  large distance $r_{12} \gg |q_1|,|q_2|$.
Our goal is to examine what is happening
 with their interaction if there are other
instantons in the vicinity of either instanton 1 or 2.
We know that the most important role in the problem
play the fermionic zero-modes.
We assumed in the previous consideration that the scalar field
is weak enough, it does not seriously change the behavior of the zero-modes.
The basis for this assumption provides the inequality $f \phi \ll 1/\rho$.
When one considers very large separation of instantons, then the zero-mode
is to be considered far outside of the instanton radius where even
the weak scalar field can drastically change the zero-mode behavior.
Note that in the latter consideration we continue to call the zero-modes
those functions which originate from  zero-modes of the pure gauge
theory. The scalar condensate influence upon them, shifting their
eigenvalues from the zero value but this shift is supposed be small.

Calculating the fermionic determinant Eq.(\ref{detdet})
one is to face  an eigenvalue problem Eq.(\ref{eigen}),
$\left( \gamma \nabla + V \right ) \psi = \varepsilon \psi$.
It is convenient for our purposes to re-wright it as
\begin{equation}\label{eivp}
\psi(x) 
= -\int G(x-x',\varepsilon)\gamma_\mu A_\mu(x') \psi(x')d^4x'~,
\end{equation}
 where $A_\mu(x)= -i A_\mu^a(x) \tau^a/2$,
and $G(x,\varepsilon)$ is the Green function whose kernel $G(\varepsilon)$ is
\begin{equation}\label{Glam}
G(\varepsilon)= \frac{1}{ \gamma \partial + V -\varepsilon}=
\frac{\gamma \partial -V_0 (1+\gamma_5)/2+\varepsilon}{\partial^2+
\varepsilon (V_0-\varepsilon)}~,
\end{equation}
where $V_0=f\phi \vec T\vec U$.
Eq.(\ref{eivp}) is convenient for the asymptotic expansion because
in the singular gauge the vector-potential $A(x')$ is localized on the 
instanton. 
Consider the zero  mode $\psi(x)=\psi_s(x)$ localized on the instanton $s$.
Then the integration in Eq.(\ref{eivp})
is also localized in the vicinity
of the instanton $s$. As a result we find from 
Eq.(\ref{eivp}) the asymptotic
\begin{equation}\label{psias}
\psi_s(x) \rightarrow 
G(x-y_s,\varepsilon) \lambda_s~, ~~|x-y_s|\rightarrow \infty~,~
\end{equation}
where $\lambda_s=- \int \gamma_\mu A_\mu(x')\psi_s(x')d^4 x'$.
It is easy to find the constant $\lambda_s$. Really, we assume that the
scalar field $f \phi$ is so weak that the zero-mode
is not influenced by the condensate in the instanton vicinity. Therefore
$\lambda_s$ may be found using the pure zero-modes, non-affected by the 
scalar condensate. The pure zero-mode $\Psi_s(x)$
satisfies the asymptotic condition similar
to Eq.(\ref{psias}) with the Green function being
$G_0(x)=\gamma\partial/\partial^2 = 
\gamma x/(2\pi^2x^4)$.
Comparing this with Eqs.(\ref{Psixan}),(\ref{Lgas}) one finds
that $\lambda_s$ is the left-hand Dirac spinor
$\lambda_s = \lambda_{s,A}^L(\alpha,n) =2\pi q_{s,nn'} 
\epsilon_{n'\alpha}\omega_A$ .

We see from Eq.(\ref{Glam}) that the asymptotic behavior very strongly
depends on the eigenvalue $\varepsilon$. For the pure zero-mode 
$\varepsilon=0$
there is only the possibility for the power-type decrease of the zero-mode
with separation.
Even the small  shift of the eigenvalue from the zero value
changes the situation drastically.
If $\varepsilon \ne 0$, then
there appears the constant 
$\varepsilon (V_0-\varepsilon)$ in the denominator 
in the Green function Eq.(\ref{Glam})
resulting in the exponential decrease of the zero-mode Eq.(\ref{eivp}).

This consideration justifies the results of Section IV for the two-instanton
problem. Really, in this case
there certainly appear the nonzero eigenvalues, but their magnitude decreases
with separation  as $\sim 1/r^4$. As a result one can 
neglect $\varepsilon \approx 0$ in the denominator of the Green
function, thus resulting in the power type decrease of the zero-mode.

The situation  changes if there are other instantons.
Let for example
there is the instanton $1^{'}$ in the vicinity of the instanton 1 and the 
instanton
$2^{'}$ in the vicinity of the instanton 2. 
Let their separations satisfy  $ r_{11^{'}}, r_{22{'}}  \ll 
r_{12}$.
Then the eigenvalues of the zero-modes centered on the instantons $1,1^{'}$
are influenced  mainly by their mutual interaction, being independent 
on the large distance $r_{12}$. The same happens with eigenvalues of
the zero-modes centered on  the instantons $2,2^{'}$.
This  results  in the exponential decrease of the zero-modes. 
The zero modes centered
on instantons $1,1^{'}$ or on $2,2^{'}$ decrease as
\begin{equation}\label{decr}
\psi_s \sim \exp (- M_s |x-x_s|),~~s=1,2~. 
\end{equation}
Here $M_s =(\varepsilon_s^2 -  \varepsilon_s V_0)^{1/2}$.
According to Eq.(\ref{Glam}) it depends on the splitting 
$\varepsilon_s=\varepsilon_{s\pm}$ 
which  can be found from Eqs.(\ref{epspm}),(\ref{Wfin}),
$\varepsilon_{s\pm}\approx\pm f \phi |q_s| |q_{s^{'}}| 
\sin \gamma_{s s^{'}}/(2 r^2_{ss^{'}})$. $M_s$ depends  as well on the 
eigenvalues $f\phi/4, -3 f\phi/4$ of the operator $V_0=f\phi \vec T \vec U$.
The exponential decrease of the zero modes with separation 
results in the decrease
of the matrix element $\vec W_{12}$
responsible for the interaction between instantons 1 and 2. 
This consideration shows that the interaction of the
instantons $1,2$ can be described by the power law Eq.(\ref{SFdilgas}) only
if 
\begin{equation}\label{rmax}
r_{12}< r_{\rm max}=\frac{1}{f\phi}~.
\end{equation}
We estimate here the splitting of the eigenvalues of instantons
$1,1^{'}$ 
and $2,2^{'}$ as $ |\varepsilon_{s\pm}| \sim f\phi$. 
% (Note, that we neglected in the right-hand side
% of Eq.(\ref{rmax}) the coefficient which is less then unity.
% This seems to be a reasonable procedure, because in the vacuum
% of instantons one should expect that each one is surrounded by ``several''
% others. This ``several'' should compensate for the coefficient.) 

We conclude, if $r_{12}>r_{\rm max}$ then the interaction between
instantons decreases exponentially with separation $r_{12}$.

\section{The mean field approximation}

Let us examine the possibility of the phase transition into the state
with polarized instantons in the mean field approximation. 
Consider the ensemble of instantons. According to Eq.(\ref{SFdilgas}) their
interaction is described by the action
\begin{equation}\label{sumin}
S_F = \zeta^2 {\sum_{s\ne t}}'  \frac{ |q_s|^2 |q_t|^2}
{r_{st}^4} \left(1-(n_{s,\mu}n_{t,\mu})^2\right)~.
\end{equation}
We took here into account that $\sin^2 \gamma_{st}=1-(n_{s,\mu}n_{t,\mu})^2$,
where $n_{s,\mu}=q_{s,\mu}/|q_s|$ is the unit vector characterizing
the orientation of the instanton $s$.
It is shown in Section V.A that the interaction of instantons is suppressed
for small separation $r_{st}<|q_s|+|q_t|$. In Section V.B
it is shown that the interaction is suppressed  for 
 large separation $r_{st}>r_{\rm max}$ as well, here $r_{\rm max}$ is 
estimated
in Eq.(\ref{rmax}) $r_{\rm max} \sim 1/(f\phi)$.
The suppression of the interaction in the asymptotic regions is taken 
into account in Eq.(\ref{sumin})
by neglecting the contribution of the 
suppressed regions. The prime over summation in
Eq.(\ref{sumin}) reminds  one of this fact.

The corresponding statistical sum for the  ensemble of instantons is
\begin{equation}\label{zst}
Z= \sum w \exp (-S_F)~,
\end{equation}
where summation runs over the number of instantons, their
positions, radiuses and orientations. 
The summation includes also the fluctuations above the pure instanton picture.
The function $w$ in  Eq.(\ref{zst})
is the probability to find the noninteracting ensemble of instantons
in some particular state.

In the  mean field approximation
one instanton is to be considered in the effective  field created by the other
instantons. 
The corresponding action is to be obtained  averaging the action (\ref{sumin})
over the positions, radiuses and orientations of other instantons. Moreover,
we are mainly interested in one and only one
degree of freedom for the considered instanton,
its orientation. Therefore it is natural to fulfill the 
averaging  over the radius of the considered
instanton as well. As a result
one finds the effective action 
in the mean-field approximation
as the function of the instanton orientation $n$
\begin{equation}\label{amf}
S_{\rm mf}(n) =  A n_\mu n_\nu <n_\mu n_\nu>~,~~{\rm where}~~A=
2 \zeta^2 \frac{ \rho_0^4}{R^4}~.
\end{equation}
Here $\rho_0^2= <\rho^2>$ is the averaged square of the instanton radius.
Remember that there exists the scalar condensate which makes this radius 
to be finite, $\rho_0 \sim 1/M_V \sim 1/(g \phi)$, see Eq.(\ref{mV}).
The quantity $<n_\mu n_\nu>$ describes the averaged orientation of the
instantons in the vacuum. 
For non-polarized vacuum, when there is no preferred
orientation, $<n_\mu n_\nu>=  \delta_{\mu\nu}/4$.
The factor $1/R^4$ describes the separation of the instanton from other ones
summed over their positions
\begin{equation}\label{R4}
\frac{1}{R^4} = N \int d^4 r \frac{1}{r^4} =
2 \pi^2 N \ln \frac{r_{\rm max}}{r_{\rm min}}~,
\end{equation}
where $N$ is the instanton concentration. We can put
$r_{\rm min}\approx 2 \rho_0\sim 1/(g \phi)$, 
because for shorter separation 
instantons do not interact, while $r_{\rm max}$ according to Eq.(\ref{rmax})
is $r_{\rm max}\approx 1/(f\phi)$.  
Thus $\ln (r_{\rm max}/r_{\rm min})\approx \ln(g/f)$.
 Using Eq.(\ref{R4}) one can present the parameter
$A$ in the form  
\begin{equation}\label{A8}
A=8\zeta^2 v_0 N L \equiv 8 B~,~~~{\rm where} ~~~B=\zeta^2 v_0 N L~. 
\end{equation}
Here $v_0=\pi^2\rho_0^4/2$ is the volume occupied by the instanton with
the radius $\rho_0$ and $L=\ln (r_{\rm max}/r_{\rm min})\sim
\ln (g/f)$.

The probability to find the instanton with orientation $n$
in the mean-field approximation is
$w_{\rm mf}(n)= \exp S_{\rm mf}(n)$.
The self-consistency condition for the averaged orientation $<n_\mu n_\nu>$
is
\begin{equation}\label{scc}
<n_\mu n_\nu> =\frac{\int n_\mu n_\nu \exp (A <n_\sigma n_\tau>
n_\sigma n_\tau)d n}
{\int \exp (A <n_\sigma n_\tau>n_\sigma n_\tau)d n}~.
\end{equation}
The integration here covers the sphere $n_\mu n_\mu =1$.
The fact of polarization of instantons means that there is at least
one eigenvalue of the matrix $<n_\mu n_\nu >$ which exceeds $1/4$.
In order to find such an eigenvalue
it is sufficient to  use for this matrix the anzats
\begin{equation}\label{anz}
< n_\mu n_\nu > = \alpha^2 \delta_{\mu4}\delta_{\nu4}+\frac{1-\alpha^2}{4}
\delta_{\mu\nu}~,
\end{equation}
in which it is supposed that the possible preferred orientation takes
place along the $4$-axes. Substituting Eq.(\ref{anz}) into Eq.(\ref{scc})
one finds the equation for $\alpha^2$
\begin{equation}\label{alf}
\frac{z}{8 B} = f(z), ~~~{\rm where}~~~z=8 B \alpha^2~,
\end{equation}
and the function $f(z)$ is
\begin{equation}\label{f(z)}
f(z)= \frac{4 \int n_4^2 \exp ( z n_4^2) dn}
{3\int \exp (z n_4^2)dn} -\frac{1}{3} = \frac{1}{3}
\frac{ I_1(z/2)-I_2(z/2) }{ I_0(z/2)-I_1~(z/2) }~.
\end{equation}
Here $I_m(x),~m=0,1,2$ are the modified Bessel functions.
The function $f(z)$ has the properties:
$f(z)\rightarrow 1,~ z \rightarrow \infty,~ 
f'(0)= 1/12 \approx 0.083< \max (f(z)/z) \approx 0.110$.
The later inequality indicates that $f''(z)$ changes the sigh.
Fig.2  shows  $f(z)$
 as well as the left-hand side of 
Eq.(\ref{alf}) for different parameters $B$. The nontrivial solutions
of Eq.(\ref{alf}) exist if 
\begin{equation}\label{cri}
B= \zeta^2 v_0 N L \ge B_c = 1.14~,
\end{equation}
It is clear that $v_0 N$ is to be small, $v_0 N<1$, 
otherwise the instanton overlapping would diminish their interaction.
It is natural to put aside the peculiar
possibility  that $\ln (g/f)$ may be considered as a large parameter.
Then the only way to satisfy
inequality in Eq.(\ref{cri}) is to have $\zeta>1$. 
It is important that this
region of $\zeta$ can be considered in the framework of the 
formalism presented above. 
Really, the only place where the perturbation theory over $\zeta$
was used is the evaluation of  the action Eq.(\ref{SFdilgas}).  
One should use
here the perturbation theory over the parameter $\zeta^2 \rho_1^2 \rho_2^2 
\sin^2 \gamma_{12}/r_{12}^4<1$, which can be small even for large
$\zeta>1$.
  
We see from Fig.2 that
if $B > B_c'=3/2$ then the nontrivial solution is unique. In the region
$B_c < B < B_c'$ there are two non-zero solutions. One of them is unstable
in the following sense. The fluctuation of the instanton
polarization in the partition
function described by the right-hand side of Eq.(\ref{scc})  results in the 
stronger fluctuation of the averaged polarization  in the
left-hand side. The existence of nontrivial solutions means that
there appears the state with polarized instantons. The fact that 
there are two solutions in the vicinity of the critical value $B_c$
indicates that the nontrivial phase appears as a result of  the first-order
phase transition.

Notice that the action Eq.(\ref{sumin}) 
depends on the scalar products of the vectors $n_{s,\mu}$. This feature
 makes it look similar to  the classical
Heisenberg model for ferromagnetics. The only distinction is the second power
of the scalar product $(n_{s,\mu} n_{t,\mu})^2$ in Eq.(\ref{sumin}). 
There is a clear
topological reason for this peculiarity. An instanton describes the 
gauge field which belongs to the adjoint representation of the gauge
$SU(2)$ group. The different $SU(2)$ matrixes $\pm n_\mu \tau^+_\mu$
have the same representatives in the adjoint
representation. In this sense $n_\mu \equiv -n_\mu$. Therefore
all the quantities describing the instanton can depend only
on even powers of $n_\mu$.
In the mean field approximation this  second power $(n_{s,\mu} n_{t,\mu})^2$
strongly manifests itself.
It results in the change of sign of $f''(z)$ which gives
two nontrivial solutions for $B_c<B<B_c'$, and consequently the first
order phase transition.

Let us discuss shortly the most important assumptions which are made in 
this section.\\ 
1.We  consider the contribution of the instantons to the
statistical sum. The quantum fluctuations in the vacuum above the instantons
are divided into two parts. One of them is the 
one-loop contribution of the fermions. 
These particular fluctuations are of grate importance because they provide
the interaction between instanton making  
their identical orientation  be more probable. They are
taken into account in Eq.(\ref{zst}) explicitly 
through the action $S_F$. 
Notice that the role of these fluctuations is enhanced by the parameter
$\zeta >1$.
All the rest fluctuations present themselves in 
Eq.(\ref{zst}) implicitly in the function $w$.
In the mean field approximation all these fluctuations manifest themselves
through the averaged values $v_0,N,L$.\\
2.We assumed that the picture based on the dilute gas approximation
for the instantons is valid. There are two reasons in favor of this 
assumption.
First, there is  the scalar condensate  in the model
which suppresses the instantons of large radius.
Second, if there is an overlapping between some  instantons violating
 the picture of the dilute gas approximation, 
then still we know that the interaction of these overlapping 
instantons is suppressed. Therefore only the configurations with 
non-overlapping instantons should  play an important role.\\
3. The instantons  are taken into account explicitly, the antiinstantons 
are neglected. The reason in favor of  this approach
gives a fact that in the considered model the instantons
interact and the antiinstantons do not. There is however an interesting 
problem: there exists the instanton-antiinstanton interaction  
 \cite{CDG}. It is necessary therefore to examine the role of this interaction
in the considered vacuum. This problem remains outside the scope of the
present paper.\\
4.The interaction of instantons is considered as a sum of two-bodies
interactions. The virial corrections caused by simultaneous
interaction of three or  more instantons
are not considered. This approximation is consistent with the dilute
gas approximation.\\ 
5.The mean field approximation for the ensemble of instantons 
is implemented in the most simple way.
Only one degree of freedom - the polarization of instantons
is taken into account in the effective action.
All other degrees of freedom  are averaged out of the action.

In conclusion, the mean field approximation confirms
the existence of the vacuum  with polarized instantons. 
The transition to the polarized state in this approximation
is the first-order phase transition.

\section{The models with $ SU(2) \times SU(2)\subset G$ gauge symmetry}
%************************************************************
Consider the gauge group $G$ which is big enough to include the product of
two $SU(2)$ groups: $G_1\times G_2 \subset  G$, where $G_1 = G_2 =SU(2)$.
Then the idea of the model developed in Section II may be presented in the 
other  useful way.
It is sufficient to consider 
in this case one generation of fermions which belongs to
the fundamental representation
of both  $G_1$ and $G_2$. Let $\Phi(x)$ be a scalar field in the vector 
representations of $G_1$ and $G_2$
\begin{equation}\label{PhiTT}
\Phi(x)=\Phi_{a,b}(x) T^a_1 T^b_2~,
\end{equation}
here $\vec T_1, \vec T_2$ are the generators of $G_1$ and $G_2$. 
Let us introduce the scalar-fermion interaction described by
the Lagrangian
\begin{equation}\label{PPP}
{\cal L}_{sf}(x) = f \Psi^+(x) \Phi(x) \frac{1-\gamma_5}{2} \Psi(x)~.
\end{equation}
Suppose that the field
$\Phi(x)$ develops the condensate
\begin{equation}\label{Pcond}
(\Phi(x))_{cond} =\phi \vec T_1 \vec T_2~.
\end{equation}
Then there appears the field $V$ which influence upon the right-hand fermions
\begin{equation}\label{Vn}
V = f \phi \vec T_1 \vec T_2 \frac{1-\gamma_5}{2}~.
\end{equation}
Consider now several instantons whose gauge field is transformed under $G_1$.
Then we can identify  vector $\vec T_1$ in Eq.(\ref{Vn}) with $\vec T$ in
Eq.(\ref{V}) and vector  $\vec T_2$  with $\vec U$. 
The fact that $\vec T_2$
generates the gauge group does not play a role here because the instantons
under consideration belong to $G_1$, they are not transformed under $G_2$.
Therefore the fermion determinant for the considered case is identical
to the determinant discussed in the previous sections, and
Eqs.(\ref{detk=2}),
%  (ref{detk=3}),
(\ref{perth}),(\ref{detk=2g}) remain valid
for the considered model as well.
 As we know there appears the 
 interaction between instantons with tendency to make their orientation be
identical, see (\ref{SFdilgas}).
The same is true for the instantons belonging to 
$G_2$.

Up to this point there is the close similarity beween the considered model and
the one discussed previously in Section II. 
But there is an important distinction. The
scalar condensate (\ref{Pcond}) is invariant under the $SU(2)$ gauge
group generated by $\vec T =\vec T_1 +\vec T_2$. 
Therefore the three vector fields
transformed under this group acquire no mass from the scalar condensate
\begin{equation}
M_V =0~.
\end{equation}
This is in contrast to the model of Section II where all the gauge fields
possess the mass (\ref{mV}).

\section{Conclusion}
%*****************************************************
It is shown that the considered models provide the strong interaction between
instantons making their identical orientation to be more preferable.
At the same time there is no such interaction between the antiinstantons.
One can easily reverse the situation changing the sign in front of $\gamma_5$
in Eq.(\ref{V}) or in Eq.(\ref{PPP}). 
Then antiinstantons interact and instantons do not. 
The interaction appears  from 
one-fermion-loop correction to the  gauge action. 
%Though we did not consider 
% the contribution of the loops of scalars and
%gauge fields, 
%we can rely upon the obtained result if the
%specific conditions are fulfilled. 
%Namely, consider the case when  both the scalar
%condensate and the mass of the fermion is small, $ \phi\rightarrow 0,
%~m \rightarrow 0$, 
%keeping their ratio as a constant $\zeta = f \phi/(2m) = const$. 
%In this limit
%the obtained result survives because it depends on $\zeta$. In contrast,
%all the other one-loop corrections to the action are reduced to the ones
%in a pure gauge theory
%with zero value of the scalar condensate.
%They are  recognized to give no  interaction between the instantons.
%Therefore the found correction to the action dominates in this limit.

We considered two models. One based upon the $SU(2)$ gauge group.
The other model is based upon 
larger gauge group $G_1\times G_2 \subset G,~G_1=G_2=SU(2)$.
These two models agree
in supplying the instantons with the interaction. The advantage
of the model based upon the wider group is the fact that it preserves
the massless gauge fields. These fields are transformed under the gauge group
$SU(2)$ overlapping with the groups $G_1,~G_2$ in which the interaction
of instantons takes place. 
If the condensate of polarized instantons is developed in
$G_1$ and (or) $G_2$ then we will enjoy the possibility
 to consider the interaction of the
massless gauge fields with that condensate. This is exactly what is necessary
for the construction of Ref.\cite{IAP}.

It is shown  that the mean field approximation confirms the
existence of the phase with polarized instantons
which is separated from the non-polarized phase by the first-order phase
transition. 
 
The obtained results give a hope
that  there exists the state with polarized instantons in gauge theory.

\acknowledgments

I am thankful to V.L.Chernyak,
V.V.Flambaum, C.J.Hamer, I.B.Khriplovich, and O.P.Sushkov
for critical comments, 
 G.F.Gribakin for helpful remarks,  A.Amann for assistance in numerical 
calculations, and L.S.Kuchieva for help
in preparation of the manuscript.
Participation in  the Workshop on Non-Perturbative Methods
in Field Theory held  by Australian
 National Center for Theoretical Physics,
Canberra 1995,  was helpful in finishing this paper.
The financial support of Australian Research Council is acknowledged.

\section{Appendix A}
%****************************************************************
In this paper the following definitions for isospinor and spinor matrixes 
are used 
\begin{displaymath}
\tau^{\pm}_\mu=( \pm i\vec \tau, 1),~~ \sigma^{\pm}_\mu=
(\pm i \vec \sigma, 1)~.
 \end{displaymath}
%The normalization differs from VZNS. In VZNS 
%  \tau^{\pm}_\mu=( \vec \tau,1),\mp i). Therefore 
%  \tau^{+}_{\mu, here}= i\tau^{+}_{\mu,VZNS},
%  \tau^{-}_{\mu, here}=-i\tau^{+}_{\mu,VZNS}
The t'Hooft symbols are
\begin{eqnarray}\nonumber
&&\eta^a_{\mu\nu}=\epsilon_{\mu\nu a4}+\delta_{a\mu}\delta_{4\nu}-
\delta_{a\nu}\delta_{4\mu}~,
\\ \nonumber
&&\bar \eta^a_{\mu\nu}=\epsilon_{\mu\nu a4}-\delta_{a\mu}\delta_{4\nu}+
\delta_{a\nu}\delta_{4\mu}~.
 \end{eqnarray}
The  Dirac matrixes $\gamma_\mu$ in the Eucledean space  are 
$\gamma_\mu=(-i\vec \gamma^M,\gamma^M_0)$,  where $ \gamma^M_\mu$
are the usual Dirac matrixes in Minkowsky space. The Eucledean matrixes 
satisfy
 the condition
$\{\gamma_\mu, \gamma_\nu \}=2 \delta_{\mu\nu}$.
In the spinor representation they have the form
\begin{eqnarray}\label{dirm}
\gamma_\mu=
\left(  \begin{array}{cc}
0& \sigma^+_\mu\\ \nonumber
\sigma^-_\mu& 0
\end {array}\right),~
\gamma_5=\left(  \begin{array}{cc}
-1 & 0\\ \nonumber
0 & 1
\end {array}\right)~.
\end{eqnarray}

Consider the  $k$-instanton general  solution \cite{ADHM}, 
for the revew see   \cite{Pras}.
Let us introduce a quaternion as $q=q_\mu \tau^+_\mu$, where
 $q_\mu, \mu=1, \cdots,4$ is an arbitrary vector. In this notation 
$q^+=q_\mu \tau^-_\mu$.
Consider the $(k+1)\times k$ matrix $M_{s,t}(x),
s=1,\cdots,k+1,~t=1,\cdots,k$ which has quaternionic matrix elements.  
$M(x)$ is supposed to be a linear function of the quaternion of coordinates 
$x=x_\mu \tau^+_\mu$
\begin{displaymath}
M(x)=B-Cx~,
\end{displaymath}
where $B$ and $C$ are $x$-independent $(k+1)\times k$ quaternionic matrixes.
They must be chosen so that the condition
\begin{equation}\label{Rx}
M^+(x)M(x)=R(x)
\end{equation}
be fulfilled for any $x$. 
Here $R(x)$ is a non-degenerate $k\times k$ real matrix.
The matrix $C$ can be chosen to be
\begin{equation}\label{Cmatr}
C_{1t}=0,~ ~C_{1+s,t}=\delta_{st}, ~s,t=1,\cdots,k~.
 \end{equation}
Having $M(x)$ one can find the $k+1$ quaternionic vector $N(x)$ which
satisfies the equations
\begin{displaymath}
M^+(x)N(x)=0,~N^+(x)N(x)=1~.
\end{displaymath}
Then the vector-potential defined in the  quaternion representation as
 \begin{displaymath}
A_\mu(x)=-i A_\mu^a(x)\tau^a/2=N^+(x)\partial_\mu N(x)
\end{displaymath}
results in  the general self-dual gauge field for $k$ instantons
$F_{\mu \nu}(x)=-i F^a_{\mu\nu}(x)\tau^a/2 =[\nabla_\mu,\nabla_\nu]$,
where $\nabla_\mu =\partial_\mu+A^+_\mu(x)$.
For single instanton 
\begin{displaymath}
M(x) = 
\left(\begin{array}{c}
q \\ y-x \end{array}\right),~N(x)=\frac{1}{\rho_0(x)}
\left(\begin{array}{c}1\\ \frac{x-y}
{|x-y|^2}q^+\end{array}\right)u~,
\end{displaymath}
where $\rho^2_0(x)=1+\rho^2/|x-y|^2$.
The quaternion $q$ describes the instanton orientation,
which may be considered as a unit four-dimensional vector
$n=q/|q|$,
and the instanton
 radius $\rho=|q|$, the position of the instanton is given by the 
quaternion $y$. The quaternion $u$ satisfying condition $|u|^2=1$
describes the freedom of the 
 gauge transformations. The condition $u=1$ corresponds
to the singular gauge \cite{tH} in which
the  vector potential has the form 
\begin{equation}\label{1ins}
A_\mu(x)=-\frac{i}{2\rho_0(x)}
\eta^a_{\mu\nu}\frac{q\tau^a q^+(x-y)_\nu}{|x-y|^4}~.
\end{equation}
Similar simple formulas are valid for several instantons
if they are well-separated
$\rho\ll r$, where  $\rho$ is the radius of an instanton and $r$ is a 
separation
of this instanton from the others. Then the dilute gas approximation 
considered in Ref.\cite{ChW} is valid.  
In this approximation the first row of the 
matrix $M(x)$ is given by $k$ constant quaternions
\begin{equation}\label{M1t}
M_{1,t}(x)=q_t, ~t=1,\cdots,k~.
\end{equation}
while the remaining matrix elements  may be approximated as
\begin{equation}\label{M1st}
M_{1+s,t}(x) = \delta_{st}(y_t-x),~s,t=1,\cdots,k~.
\end{equation}
Here $q_t$ describes the radius and orientation of the $t$-th instanton,
$y_t$ is the position of this instanton.
The components of the vector $N(x)$ are
\begin{eqnarray}\label{N1}
&&N_1(x)=\frac{1}{\sqrt{\rho_0(x)}}u~,
\\ \label{N1t}
&&N_{1+t}(x)=\frac{1}{\sqrt{\rho_0(x)}}\frac{x-y_t}{|x-y_t|^2}q^+_t u,~
t=1,\cdots,k,
\end{eqnarray}
where 
\begin{equation}\label{rho0}
\rho_0(x)=1+\sum_{t=1}^{k} \frac{|q_t|^2}{ |x-y_t|^2}~.
\end{equation}
The vector potential for $k$ well separated instantons in the singular gauge
$u=1$ has the form 
\begin{equation}\label{Adg}
A_\mu(x)=-\frac{i}{2 \rho_0(x)}
\eta^a_{\mu\nu}\sum_{t=1}^k \frac{q_t \tau^a q_t^+
(x-y_t)_\nu}{|x-y_t|^4}~.
\end{equation}
There is a particular case \cite{CF},\cite{tH1},\cite{Wi} which is
interesting for our purposes:
the  identical orientation of the instantons
\begin{equation}\label{idor}
q_t=\rho_t w~.
\end{equation}
Here $\rho_t$ is the real radius of $t$-th instanton and the quaternion
$w,~w^+w=1$, describes the common orientation of all the instantons.
It is known \cite{ChW} that in this case 
Eqs.(\ref{M1t}) -- (\ref{Adg}) are valid for any separation
between instantons.

\section{Appendix B. The fermionic zero-modes}
%*******************************************************************
The purpose of this section is to evaluate the convenient form for 
fermionic zero-modes in the field of several instantons.
The projection operator $P$ onto the 
space of fermionic zero-modes was obtained in \cite{Brown}
in the following form
\begin{equation}\label{Pdef}
P=\left( 1-(\gamma \nabla)\frac{1}{\nabla^2}(\gamma \nabla)\right)
\frac{1-\gamma_5}{2}~.
\end{equation}
The following simple arguments can be used to verify it.
It is easy to check out directly
that Eqs.(\ref{P1})
$P^2=P , ~( \gamma \nabla) P=0$ are fulfilled
and therefore $P$ gives  a projection into the  space of zero-modes.
In order to prove that it gives the projection on the hole  space of  
zero-modes 
it is sufficient to show that $Sp(P)$ is equal to the number of zero-modes.
It follows from (\ref{Pdef}) that
\begin{displaymath}
\mbox{Sp}(P) = 
\mbox{Sp} \left(  \frac{1-\gamma_5}{2}-(\gamma \nabla)^2
\frac{1}{\nabla^2}\frac{1+\gamma_5}{2} \right)~,
\end{displaymath}
and therefore
\begin{equation}\label{numzero}
\mbox{Sp}(P)=\mbox{Sp}(-\gamma_5)~.
\end{equation}
Every  zero-mode gives a unity contribution to the right-hand side.
The nonzero-modes do not contribute to it.
Indeed, for any nonzero-mode 
$\psi_\lambda(x)$  satisfying the equation
\begin{equation}\nonumber
-i(\gamma \nabla) \psi_\lambda (x)=\lambda \psi_\lambda (x),
\end{equation}
 with $|\lambda | > 0$ the function
$\psi_{-\lambda}(x) = 
\gamma_5 \psi_{\lambda}(x)$ 
satisfies $-i(\gamma \nabla)\psi_{-\lambda}(x)=
-\lambda \psi_{-\lambda} (x) $ and is, therefore, orthogonal to 
$\psi_{\lambda}(x)$. 
Thus the right-hand
side of Eq.(\ref{numzero}) is equal to the number of zero modes. 
This concludes 
the verification of the fact that $P$  defined in (\ref{Pdef}) 
is the projection operator on the zero-modes space.

The Green function $-1/\nabla^2$ describing propagation of scalars
in  the field of several instantons was 
found in  Refs.\cite{Brown},\cite{ChW}. 
For the considered case of isospin-$1/2$  it reads 
\begin{equation}\label{Gscal}
D(x,y)=-\frac{1}{\nabla^2}=\frac{1}{4\pi^2} \frac{N^+(x)N(y)}{(x-y)^2}~.
\end{equation}

Substituting this expression into Eq.(\ref{Pdef}), after lengthy but
conventional calculations presented below, see Eqs.(\ref{dg})-(\ref{Petafin}),
it is possible to
present the kernel $P(x,y)$ of the projection operator in
the following simple form
\begin{equation}\label{Pfin}
P(x,y)=\frac{1}{2\pi^2} L^+(x)C^+C\frac{1-\gamma_5}{2}
(1-\vec \sigma \vec \tau) L(y)~.
\end{equation}
Here  $L(x)$
is defined as
\begin{equation}\label{Ldef}
L(x) =R^{-1}(x)C^+N(x)~.
\end{equation}
The quantities $M(x), N(x), B, C, R(x)$ constitute the ADHM construction 
Ref.\cite{ADHM} defined  in Section VII. Note that $L(x)$ is the 
vector with  quaternionic components
$L(x)=( L_s(x),
s=1,\cdots ,k),~ L_s(x)=L_{s,\mu}(x)\tau^+_\mu$, where $k$ is the number 
of instantons.

The kernel of the projection operator satisfies the
equation  $(\gamma \nabla)P(x,y)=0$
for arbitrary $y$ and the validity of this equation does not depend
on specific properties of $L(y)$ in the representation (\ref{Pfin}).
Therefore we can obtain the expression for a zero-mode 
wave function by replacing  $L(y)$ with arbitrary 
$x$-independent spinor-isospinor $\chi$.
If we define
\begin{equation}\label{defpsi}
\psi = \sigma^+_\mu \tau^+_\mu(1-\gamma_5) \chi~,
\end{equation}
then we find from Eq.(\ref{Pfin}) the 
wave functions of the fermionic  zero modes 
\begin{equation}\label{Psizm}
\Psi_s(x) =\frac{\sqrt{2}}{\pi} L^+_s(x) \psi,  ~s=1,\cdots,k~.
\end{equation}
It follows from Eq.(\ref{defpsi}) that $x$-independent spinor-isospinor
$\psi$ satisfies the following conditions
\begin{equation}\label{psizm}
\gamma_5\psi=-\psi, ~\vec \sigma \vec \tau \psi = -3 \psi~.
\end{equation}
Let us define 
$\alpha=1,2$ the spinor index of the right-hand
spinor and $n=1,2$ the isospinor index. Then Eqs.(\ref{psizm}) result in
\begin{equation}\label{psian}
\psi = \psi^R_{\alpha,n}=\frac{1}{\sqrt{2}}\epsilon_{n,\alpha }~,
\end{equation}
where $\psi^R$ is the right-hand spinor-isospinor
defined by this equation and
 $\epsilon_{n,\alpha}$ is the usual antisymmetric tensor with $\epsilon_
{1,2}=1$.
Using Eq.(\ref{psian}) one can present the zero-modes
given in Eq.(\ref{Psizm}) in the more
detailed form
\begin{equation}\label{Psixan}
\Psi_s(x)=\Psi_s(x,\alpha,n)=\frac{1}{\pi}\left[ L^+(x)\right] _{n,n'}
\epsilon_{n',\alpha}~.
\end{equation}
The spinor $\psi$ is normalized in Eq.(\ref{psian})
 as $\langle \psi^+|\psi\rangle =1$. This results in the usual
normalization conditions for 
the functions  $\Psi_s(x)$  
\begin{equation}\label{normPsi}
\langle \Psi_t^+| \Psi_s\rangle =\delta_{st}~.
\end{equation}
The easiest way to check out these normalization conditions 
provides the  equation $P^2=P$ for the projection operator.
For the kernel $P(x,y)$ 
(\ref{Pfin}) it reads
\begin{eqnarray}\nonumber
&&(2\pi^2)^{-2} L^+(x)C^+C\sigma_\mu^+ \tau_\mu^+ \cdot 
\\ \nonumber
&&\int d^4zL(z) L^+(z)\sigma_\nu^+ \tau_\nu^+ C^+CL(y) [(1-\gamma_5)/2]=
\\ \nonumber
&&(2\pi^2)^{-1}L^+(x)\sigma_\mu^+ \tau_\mu^+
C^+CL(y)[(1-\gamma_5)/2]~.
\end{eqnarray}
To satisfy this condition $L(x)$ must obey
\begin{eqnarray}\nonumber
&&(2\pi^2)^{-1}(\sigma^+_\mu \sigma^+_\nu) \left( \tau^+_\mu 
\int L(z)L^+(z)d^4z~ \tau^+_\nu\right) =
\\ \nonumber
&&(C^+C)^{-1}\sigma^+_\mu\tau^+_\nu~.
\end{eqnarray}
Using  identity
$(\sigma^+_\mu\sigma^+_\nu)(\tau^+_\mu q\tau^+_\nu)=
4\sigma^+_\mu  \tau^+_\mu\mbox{Re}(q)$,
which is valid for any quaternion $q$ we find
\begin{equation}\label{ReKK}
\frac{2}{\pi^2}
\mbox{Re}\int L^+_t(x)L_s(x)~d^4x= (C^+C)^{-1}_{st}=\delta_{st}~.
\end{equation}
Here we choose the matrix $C$ to satisfy identity 
$\left( C^+C \right)_{st}=\delta_{st}$,
which can  be fulfilled for general $k$-instanton solution, see (\ref{Cmatr}).
The normalization  (\ref{normPsi}) follows 
from (\ref{ReKK}).

Formula (\ref{Psizm}) is valid for one generation of fermions 
in the fundamental representation
of $SU(2)$ gauge group. Generalization to the case of several
generations of fermions is strait-forward.

The  formula becomes  much simpler if the dilute gas approximation
is valid.
It follows from  Eqs.(\ref{Rx}),(\ref{M1t}),(\ref{M1st}) that in this case
the  matrix $R(x)$  simplifies to be
\begin{equation}\label{Rst}
R_{st}(x) =  \delta_{st} \left( |x-y_t|^2 + |q_t|^2\right), 
~s,t=1,\cdots,k~.
\end{equation}
Using Eqs.(\ref{Rst}),(\ref{Cmatr}),(\ref{N1}),(\ref{N1t})
one finds the simple representation for the vector $L(x)$ (\ref{Ldef})  
in the dilute gas approximation 
\begin{equation}\label{Lgasex}
L_s(x)=
\frac{x-y_s}{
\sqrt{\rho_0 (x)}
\left( |x-y_s|^2+|q_s|^2\right) |x-y_s|^2}q^+_s u~.
\end{equation}
Eqs.(\ref{Lgasex}),(\ref{Psizm})
give the fermionic zero modes in the dilute gas approximation.
For single instanton they are reduced to the known expression \cite{tH}.

Though Eq.(\ref{Lgasex}) was evaluated in the dilute gas approximation,
it proves to be  valid for the other important case when
all the instantons have identical orientation and their
separations are arbitrary.
This follows from the comment given after Eq.(\ref{idor}).

Eq.(\ref{Lgasex}) 
may be further simplified if we are interested in the region
far outside of the cores of instantons
\begin{equation}\label{dgcond}
|q_t|^2\ll |x-y_t|^2,~t=1,\cdots,k~,
\end{equation}
where it reads
\begin{equation}\label{Lgas}
L_s(x)=\frac{x-y_s}{|x-y_s|^4}q^+_s u,~ s = 1,\cdots,k~,
\end{equation}
providing the very simple expression for the fermionic zero modes
(\ref{Psizm}).

Let us verify representation Eq.(\ref{Pfin}).
With this purpose consider Eq.(\ref{Gscal}) 
from which one finds
\begin{eqnarray}\label{dg}
\nabla_\mu D(x,y) = \frac{1}{4\pi^2} \Biggl( -2\frac{(x-y)_\mu}{(x-y)^4}
N^+(x) N(y)+
\\ \nonumber
\frac{1}{(x-y)^2} \nabla_\mu N^+(x)N(y) \Biggr) ~.
\end{eqnarray}
This gives
\begin{eqnarray}\nonumber
&&(\gamma \nabla)D(x,y)(\gamma \nabla)=
(4 \pi^2)^{-1} \gamma_\mu \gamma_\nu \cdot
\\ \nonumber
&&\bigl[ -2\frac{(x-y)_\mu}{(x-y)^4}N^+(x)N(y)+
\frac{1}{(x-y)^2}\nabla_\mu N^+(x)N(y)\bigr] \nabla_\nu~.
\end{eqnarray}
Here the last operator acts on the $y$-coordinate $\nabla_\nu=
\nabla_\nu^{(y)}$.
To simplify the formulae it is usufull to apply this operator
   to the left 
\begin{eqnarray} \label{gamDgam}
&&(\gamma \nabla)D(x,y)(\gamma \nabla)=
(4\pi^2)^{-1}\gamma_\mu\gamma_\nu \cdot
\\ \nonumber
&&\biggl[ 2 \partial _\nu ^{(y)}  \bigl( (x-y)_\mu/(x-y)^4 \bigr)  
N^+(x) N(y)+
\\ \nonumber
&&2 \bigl( (x-y)_\mu /(x-y)^4 \bigr) N^+(x) (\nabla_\mu^{(y)}N^+(y))^+ -
\\ \nonumber
&&\partial _\nu ^{(y)} \bigl(  1/(x-y)^2\bigr)   \nabla_\mu N^+(x) N(y)-
\\ \nonumber
&&\bigl( 1/(x-y)^2\bigr) 
\nabla_\mu^{(x)} N^+(x) (\nabla_\nu^{(y)}N^+(y))^+\biggr]~.
\end{eqnarray}
The first term in the square brackets above yields
 \\$ \delta (x-y)$ as it follows from the simple identity
$-4\pi^2 \delta (x-y)=
\gamma_\mu  \gamma_\nu \partial _\nu^{(y)} [ 2(x-y)_\mu /(x-y)^4] =
\gamma_\mu  \gamma_\nu  \partial _\mu  \partial _\nu [1/(x-y)^2 ] $.
Then from (\ref{gamDgam})
we find for the kernel $P(x,y)$ of the operator $P$ (\ref{Pdef}) 
\begin{eqnarray}\nonumber
&&P(x,y)= [4\pi^2 (x-y)^2]^{-1}[(1-\gamma_5)/2] \gamma _\mu \gamma _\nu  \cdot
\\ \nonumber
&& \biggl[  (2/(x-y)^2) \bigl( (x-y)_\mu N^+(x) (\nabla_\nu N^+(y))^+-
\\ \nonumber
&&(x-y)_\nu (\nabla_\mu N^+(x) ) N^+(y)\bigr) -
\\ \label{gdgs}
&&(\nabla_\mu N^+(x) )(\nabla_\nu N^+(y))^+\biggr] ~.
\end{eqnarray}
Let us calculate the terms with derivatives in the right-hand side of 
(\ref{gdgs}). Note that
for the considered self-dual field the potential may be presented
as $A_\mu (x)=N^+(x)\partial_\mu N(x)$, see Appendix C. 
Therefore $\nabla_\mu N^+(x)=
\partial_\mu N^+(x) (1-N(x) N^+(x))$. Using the equality
$1-N(x) N^+(x)=M(x)R^{-1}(x)M^+(x)$ one finds
\begin{eqnarray} \nonumber
\nabla_\mu N^+(x)=\partial_\mu N^+(x)M(x)R^{-1}(x)M^+(x)=\\ \nonumber
-N^+(x)\partial_\mu M(x)R^{-1}(x)M^+(x)=\\ \label{delN}
N^+(x)C\tau_\mu^+R^{-1}(x)M^+(x)~.
\end{eqnarray}
Using the identity $M^+(x)=M^+(y)-(x-y)^+C^+ $ one finds from (\ref{delN})
\begin{equation}\label{delNN}
\nabla_\mu N^+(x)N(y)=-N^+(x)CR^{-1}(x)\tau^+_\mu (x-y)^+ C^+N(y)
\end{equation}
because $M^+(y)N(y)=0$. 
From (\ref{delNN}) one finds
\begin{eqnarray}\label{xyNdN}
&&(x-y)_\mu N^+(x) (\nabla_\nu N^+(y))^+-
\\ \nonumber
&&(x-y)_\nu (\nabla_\mu N^+(x) ) N^+(y)=
\\ \nonumber
&&N^+(x)C \biggl[ (x-y)_\mu (x-y)_\nu \left( R^{-1}(x) +R^{-1}(y)\right) -
\\ \nonumber
&&i\tau^a\left[ (x-y)_\mu \eta _{\nu\lambda}^a R^{-1}(y)-
(x-y)_\nu \eta _{\mu\lambda}^aR^{-1}(x)\right] \cdot
\\ \nonumber
&&(x-y)_\lambda \biggr] C^+N(y)~.
\end{eqnarray}
Here the identity $\tau^+\tau^-=\delta_{\mu\nu}+i \eta_{\mu\nu}^a\tau^a$ was 
used.
Now we use Eq.(\ref{xyNdN}) to simplify the first two
terms in  square brackets in (\ref{gdgs}) and Eq.(\ref{delN}) for the 
last term in (\ref{gdgs}).
As a result we find
\begin{eqnarray} \label{Pinterm}
&&P(x,y) = 
[4\pi^2 (x-y)^2]^{-1}[(1-\gamma_5)/2] \gamma _\mu \gamma _\nu  \cdot 
 \\ \nonumber
&&L^+(x) 
\Biggl[ (1/2) (R(x) + R(y))\delta_{\mu\nu} -
\\ \nonumber
&&2i\tau^a(n_\mu n_\lambda \eta_{\nu\lambda}^aR(x)-
n_\nu n_\lambda \eta_{\mu \lambda}^a R(y)) - 
\\ \nonumber
&&\tau^+_\mu M^+(x) M(y) \tau^-_\nu \Biggr]   L(y)~.
\end{eqnarray}
Here $n_\mu$ is a unit vector 
$n_\mu = (x-y)_\mu/\sqrt{(x-y)^2}$.

Now let us use the  following identity for the Dirac matrixes
\begin{equation}\label{ggg15}
\gamma_\mu \gamma_\nu (1-\gamma_5)=(\delta_{\mu\nu}+
i\eta^a_{\mu\nu} \sigma^a) (1-\gamma_5)~.
\end{equation}
Substituting this identity into  (\ref{Pinterm}) we get two terms, one comes 
from the $\delta_{\mu\nu}$ and the other one - from  $i\eta^a_{\mu\nu} 
\sigma^a$
in the right-hand side of (\ref{ggg15})
\begin{equation}\label{deleta}
P(x,y) =P_{(\delta)}(x,y) +P_{(i\eta\sigma)}(x,y) ~.
\end{equation}
For the first term we find  
substituting $\gamma_\mu\gamma_\nu \to \delta_{\mu\nu}$ in (\ref{Pinterm})
\begin{eqnarray}\label{Pdel}
&&P_{(\delta)}(x,y)=[4\pi^2(x-y)^2]^{-1} [(1-\gamma_5)/2]  \dot
\\ \nonumber
&&L^+(x)\left[ 2 (R(x)+R(y))-\tau^+_\mu M^+(x)M(y)\tau^-_\mu \right] L(y)~.
\end{eqnarray}
With the help of the identity $\tau^+_\mu q \tau^-_\mu= 4 \mbox{Re}(q)$ 
which is valid for any quaternion $q$ one finds that
\begin{displaymath}
\tau^+_\mu M^+(x)M(y)\tau^-_\mu=2[R(x)+R(y)-(x-y)^2C^+C]~.
\end{displaymath}
Then it follows from (\ref{Pdel})
\begin{equation}\label{Pdelfin}
P_{(\delta)}(x,y)=\frac{1}{2\pi^2}\frac{1-\gamma_5}{2}L^+(x)C^+C L(y)~.
\end{equation}
The second  term in (\ref{deleta}) is to be found 
substituting $\gamma_\mu\gamma_\nu \to  i\eta^a_{\mu\nu} \sigma^a$ 
in (\ref{Pinterm})
\begin{eqnarray}\nonumber
&&P_{(i\eta\sigma)}(x,y)=[4\pi^2(x-y)^2]^{-1} [(1-\gamma_5)/2] \cdot
\\ \nonumber
&&L^+(x) i\eta^a_{\mu\nu} \sigma^a 
\{  -2i \tau^b [n_\mu n_\lambda \eta^b_{\nu\lambda} R(x)-
n_\nu n_\lambda \eta^b_{\mu\lambda} R(y)]-
\\ \nonumber
&&\tau^+_\mu M^+(x)M(y)\tau^-_\nu\} L(y)~.
\end{eqnarray}
The first  term in curly brackets  is evaluated with the help of the identity
$\eta^a_{\mu\nu}\eta^b_{\nu\lambda} n_\mu n_\lambda =-\delta_{ab}$.
The second term is simplified with the help of 
$\eta^a_{\mu\nu} \tau^+_\mu q \tau^-_\nu=4 i \tau^a \mbox{Re}(q) $ where
$q$ is an arbitrary quaternion. 
As a result one finds
\begin{equation}\label{Petafin}
P_{(i\eta\sigma)}(x,y)=-\frac{1}{2\pi^2}\frac{1-\gamma_5}{2}L^+(x)C^+C 
\sigma^a\tau^aL(y)~.
\end{equation}
Combining (\ref{deleta}),(\ref{Pdelfin}),(\ref{Petafin}) we justify  
Eq.(\ref{Pfin})
due to the identity $\sigma^+_\mu\tau^+_\mu=1-\vec \sigma \vec \tau$.

\section{Appendix C. The matrix element ${\vec W_{1,2}}$}

In this section the explicit convenient expression
for the matrix element ${\vec W}_{1,2}$ Eq.(\ref{ttaus}) is presented.
Introducing convenient notation   $W = i {\vec\tau}\cdot 
{\vec W}_{1,2}$ one can present  Eq.(\ref{ttaus}) as
\begin{eqnarray}\label{Im}
W = \frac{2}{\pi^2}{\rm Im}&&\int L^+_1(x) L_2(x) d^4 x=\\ \nonumber
\frac{1}{\pi^2}{\rm Im}
&&\int \left( L^+_1(x) L_2(x)-L^+_2(x) L_1(x)\right) d^4 x~,
\end{eqnarray}
Here the imaginary part of a quaternion $q$ is defined as 
${\rm Im}~q=(q-q^+)/2$. The second equality in 
Eq.(\ref{Im}) follows from the antisymmetry condition
(\ref{zero}) of the matrix element $\vec W_{12}=-\vec W_{21}$.
 Using  Eq.(\ref{Ldef}) 
for the functions $L_i(x),~i=1,2$ one finds
\begin{equation}\label{RR}
W = \frac{1}{\pi^2}{\rm Im}\int N^+(x) C R^{-1}(x)\Sigma R^{-1}(x)
C^+ N(x) d^4 x~,
\end{equation}
where the matrix $\Sigma$ is defined as
\begin{equation}\label{Sig}
\Sigma = 
\left(  \begin{array}{cc}
0 & 1\\
-1 & 0
\end {array}\right)~.
\end{equation}
For two instantons the matrixes $M(x),C$ 
in the ADHM solution have an explicit simple form
\begin{equation}\label{matM}
M(x) = 
\left(  \begin{array}{cc}
q_1 & q_2\\
y_1-x & b\\
b&y_2-x
\end {array}\right)~,~~~~
C = 
\left(  \begin{array}{cc}
0 & 0\\
1 & 0\\
0 & 1
\end {array}\right)~,
\end{equation}
where $q_i,~i=1,2$ are the quaternions describing the radius and orientation
of the two instantons, $y_i$ are the quaternions describing the positions
of the instantons and 
\begin{equation}\label{b}
b=\frac{y_{12}}{2 |y_{12 }|^2}\left(q_2^+q_1-q_1^+q_2\right)~,~~
y_{12}=y_1-y_2~.
\end{equation}
To simplify Eq.(\ref{RR}) it is convenient to express 
the vector $N(x)$ in terms of the matrix $M(x)$. 
With this purpose let us use
the identity
\begin{equation}\label{NM}
N(x)=a_N N_0,~~~\Pi(x)= 1-M(x)R^{-1}(x)M^+(x),
\end{equation}
where $\Pi(x)$ is the projection operator on the vector $N(x)$,
$N_0$ is any vector non-orthogonal to $N(x)$:
$N^+(x)N_0 \neq  0$, and $a_N$ 
is the normalization coefficient which satisfies
$|a_N|^2
= \left(N_0^+(1-M(x)R^{-1}(x)M^+(x)) N_0\right) ^{-1}$. It is convenient
to chose 
\begin{equation}\label{N0+}
N_0^+=( 1, 0, 0)~.
\end{equation}
Substituting  Eq.(\ref{NM}),(\ref{N0+}) in Eq.(\ref{RR}), 
using the explicit representation (\ref{matM}) for the matrix $M(x)$
and  the obvious identities $C^+N_0=0$ and
 $R^{-1}(x)\Sigma R^{-1}(x)=\Sigma / \det R(x)$ one finds
\begin{equation}\label{Wconv}
W = \frac{1}{\pi^2}{\rm Im}\int 
\frac{Q R^{-1}(x)m^+(x)\Sigma m(x) R^{-1}(x) Q^+}{\det R(x) 
\left(1-Q R^{-1}(x)Q^+\right)}~d^4 x~.
\end{equation}
Here $m(x)$ is ``the square part'' of $M(x)$:
\begin{equation}\label{mx}
m(x) = 
\left(  \begin{array}{cc}
y_1-x & b\\
b & y_2-x
\end {array}\right)~,
\end{equation}
and $Q,Q^+$ are the two-dimensional quaternions
\begin{equation}\label{QQ}
Q = (q_1,q_2)~,~~~
Q^+ = \left(  \begin{array}{cc} q_1^+ \\ q_2^+ \end {array}\right)~.
\end{equation}
Representation (\ref{Wconv}) is convenient for direct numerical
calculations as well as for asymptotic expansion in different regions.
The most simple and interesting is the case of large separation,
 $r_{12}=|y_{12}| \gg |q_1|,|q_2|$. For this case the main
contribution to the integral in Eq.(\ref{Wconv})
gives the region $| y_1-x|\sim | y_2-x| \sim r_{12}$ in which 
\begin{equation}\label{mas}
m(x) \approx m_0(x)=\left(  \begin{array}{cc} y_1-x & 0\\ 
0 & y_2 -x \end {array}\right)~,
\end{equation}
and
\begin{equation}\label{rsa}
R(x)\approx \left(  \begin{array}{cc}| y_1-x|^2 & 0\\ 
0 & |y_2 -x|^2 \end {array}\right)~.
\end{equation}
Substituting Eqs.(\ref{mas}),(\ref{rsa}) in Eq.(\ref{Wconv}) 
and taking into account that
$|Q R^{-1}(x)Q^+| \ll 1$ one finds
\begin{equation}\label{rinf}
W = 2 \frac{ {\rm Im} (q_1 q_2^+)}{r^2_{12}}~,
\end{equation}
which agrees with Eq.(\ref{Wfin})
evaluated in Section IV using the different technic.

The other important case provides the region
$r^2_{12} \gg |q_1 q_2 \sin \gamma_{12}|$   in which
the instanton separation might be not only large but small as well,
less then the instantons  radiuses, provided their orientations
are close $|\gamma_{12}|\ll 1$. In this case $b$ in Eq.(\ref{mx})
is smaller then the separation $|b| \ll r_{12}$. Therefore one can again use
the asymptotic Eq.(\ref{mas}) for $m(x)$.
In contrast the matrix $R(x)$ is to be considered more accurately,
\begin{equation}\label{mra}
R(x)_{ij}\approx q^+_i q_j+ \left( m^+_0(x)m_0(x) \right)_{ij}~,~~~~i,j=1,2~,
\end{equation}
because the separation might be less then the radiuses. 
Substituting (\ref{mas}),(\ref{mra}) in Eq.(\ref{Wconv}) one finds after
simple algebraic calculations 
\begin{equation}\label{rinfgam}
W = 2 F(\kappa_1,\kappa_2)
\frac{ {\rm Im}( q_1 q_2^+)}{r^2_{12}}~,
\end{equation}
where $\kappa_i = 2 |q_i |/r,~i=1,2$ and 
the function $F(\kappa_1,\kappa_2)$ is 
\begin{eqnarray}\label{funuv}
&&F(\kappa_1,\kappa_2)=
\frac{4}{\pi ^2} \int \left(|x|^2-1\right) \times \\ \nonumber
&&\frac{ d^4 x}{  |x-1|^2 |x+1|^2  \left[ \left(|x-1|^2+\kappa_1^2\right)
 \left( |x+1|^2+\kappa_2^2 \right) -\kappa_1^2\kappa_2^2 \right] }~.
\end{eqnarray}
The explicit form of this integral representation for 
$F(\kappa_1,\kappa_2)$ is evaluated in 
the convenient coordinate frame $y_1 =-y_2=1$. 
The integral in Eq.(\ref{funuv}) is easily reduced to the 
one-dimensional one. Its properties are discussed in detail is Section V.

\newpage
\onecolumn

\newpage
\begin{figure}[h]
\input psfig
\psfig{file=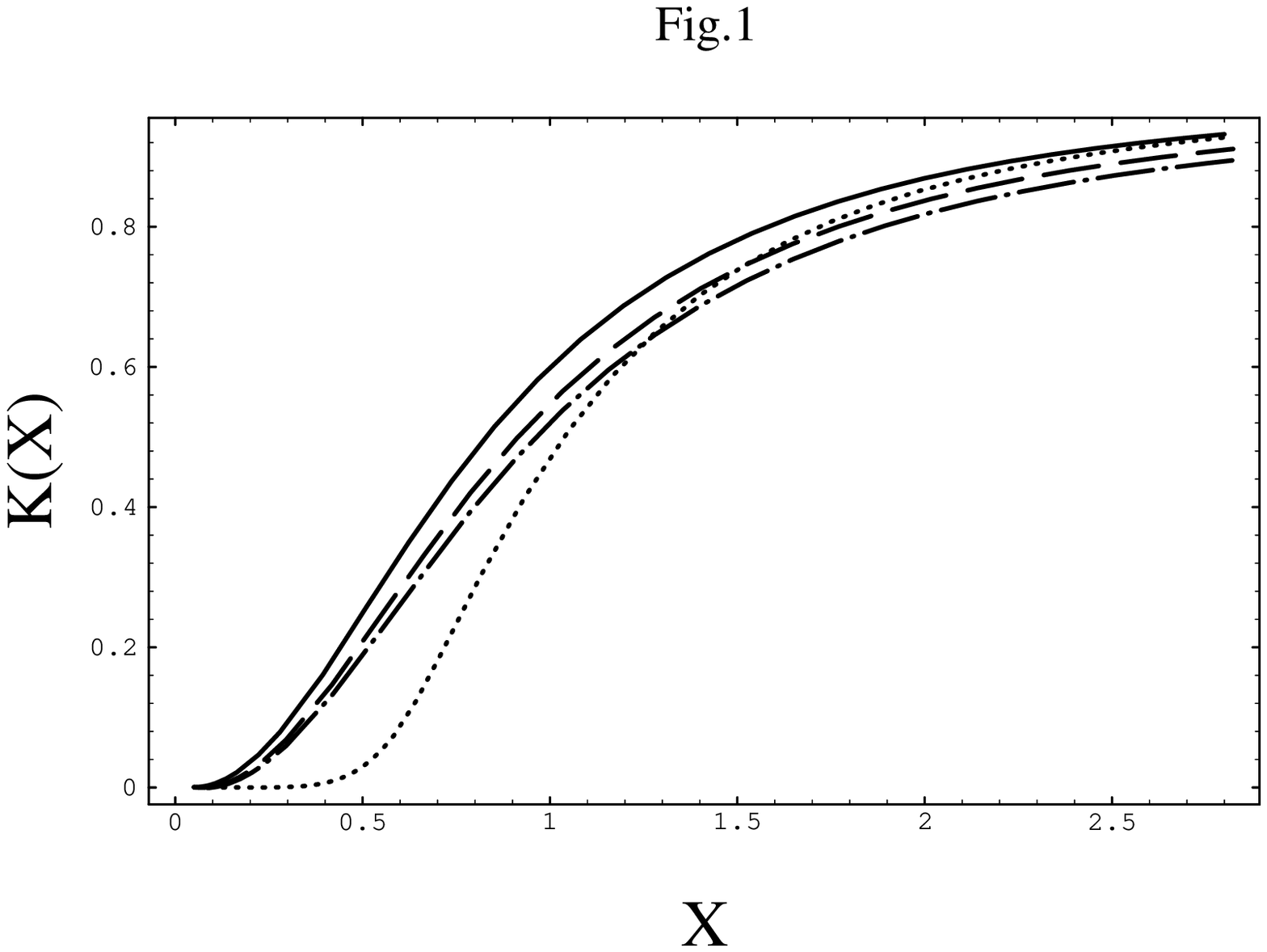, clip=}
\caption 
{ 
The interaction of instantons for small separation.
The coefficient $K(X)$ describes how strongly the interaction
of instantons deviates from the asymptotic value, see Eq.(43), 
is shown vs the relative
separation between instantons $X=r_{12}/(\rho_1+\rho_2)$. 
The different curves present the dependence of $K(X)$ on the 
ratio of instanton radiuses $\rho_1/\rho_2$ and on the angle $\gamma_{12}$
between the directions of the instanton 
orientations, $\cos \gamma_{12}=n_{1,\mu}n_{2,\mu}$.
The full curve: $\rho_1/\rho_2=1,~\gamma_{12}=0$, 
the dashed curve: $\rho_1/\rho_2=5,~\gamma_{12}=0$,
the dashed-dotted curve:
$\rho_1/\rho_2=10,~\gamma_{12}=0$,
the dotted curve: $\rho_1/\rho_2=1,~\gamma_{12}=\pi/2$.
The coefficient $K(X)$ is found
as $K(X)=F^2(\kappa_1,\kappa_2)$, see Eq.(118), for $\gamma_{12}=0$. 
For calculations with $\gamma_{12}=\pi/2$ the more general Eq.(110) is used.
The drop in $K(X)$ for small separation
indicates that in this region the instanton interaction is suppressed.}

\end{figure}

\newpage
\begin{figure}[h]
\input psfig
\psfig{file=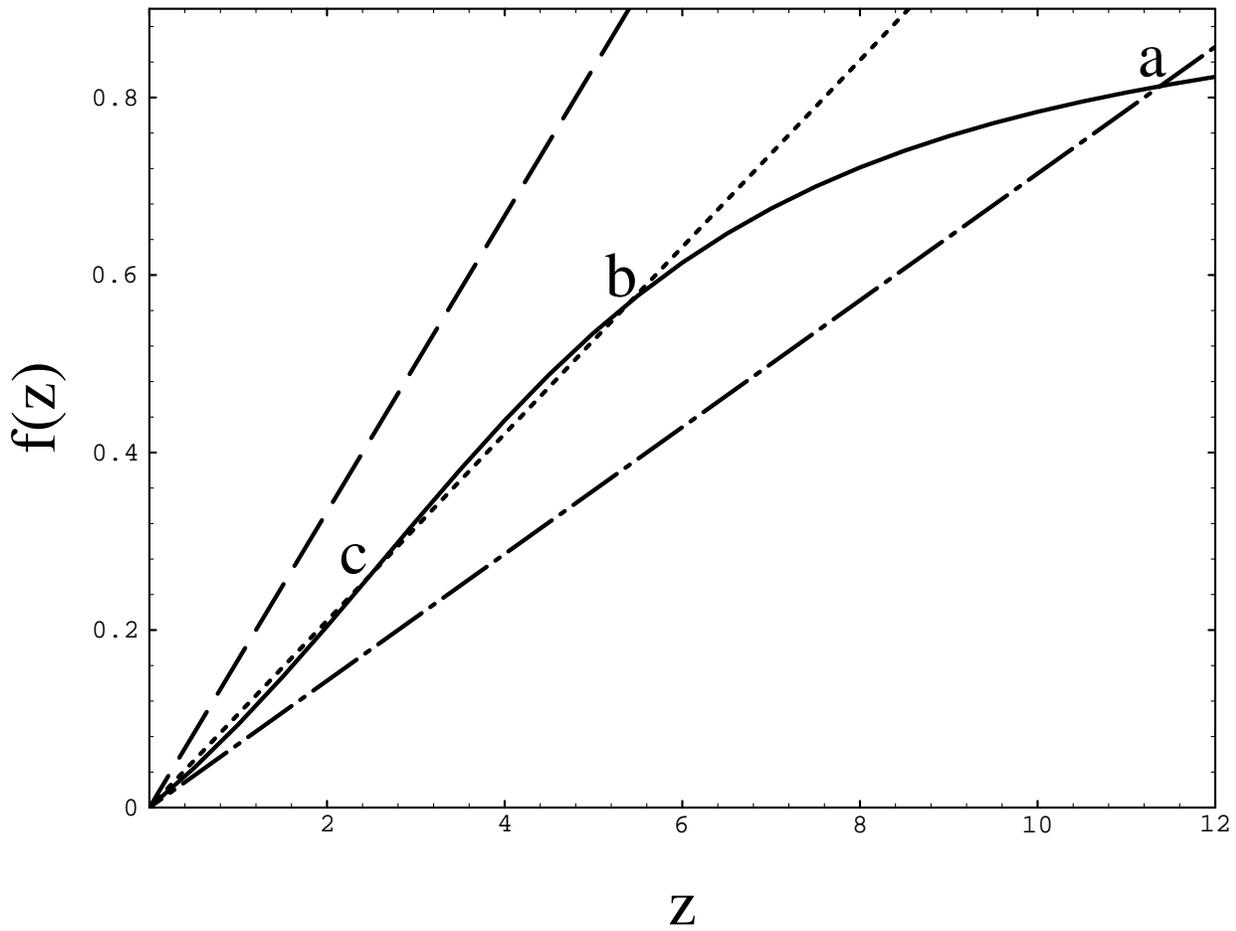, clip=}
\caption{
The   mean-field approximation.
The full curve: the right-hand side of Eq.(56) defined in Eq.(57).
The straight lines give the left-hand side of Eq.(56) for different
parameters $B$. The dashed line: $B<B_c$, the dotted line:
$B_c<B<B_c'$, the dashed-dotted curve: $B'_c<B$, where
$B_c=1.14,~B_c'=3/2$ and the three considered values of $B$ are
$B=0.75,~1.19,~1.75.$
The points ``a,b,c'' show the nontrivial solutions
in which the instantons are polarized. The point 
``a'' indicates the unique nontrivial  solution exiting for large $B$.
The points ``b,c'' are the two solutions existing for intermediate $B$.
One of the solutions, the point ``b'', is stable, the other solution,
the point ``c'', is non-stable 
to the fluctuations of the instanton polarization. 
The mean field approximation confirms the possibility of the
state with polarized instantons.}
\end{figure}

\end{document}